\documentclass[12pt,oneside]{uhthesis}
\usepackage{subfigure}
\usepackage[ruled,lined,linesnumbered,titlenumbered,algochapter,english,onelanguage]{algorithm2e}
\usepackage{amsmath}
\usepackage{amssymb}
\usepackage{amsbsy}
\usepackage{caption,booktabs}
\usepackage{float}
\usepackage{todonotes}
\usepackage{listings}
\usepackage{xcolor}
\usepackage{multicol}
\usepackage{graphicx}
\floatstyle{plaintop}
\restylefloat{table}
\SetAlgoNoEnd

\definecolor{codegreen}{rgb}{0,0.6,0}
\definecolor{codegray}{rgb}{0.5,0.5,0.5}
\definecolor{codepurple}{rgb}{0.58,0,0.82}
\definecolor{backcolour}{rgb}{0.95,0.95,0.92}

\lstdefinestyle{mystyle}{
    backgroundcolor=\color{backcolour},   
    commentstyle=\color{codegreen},
    keywordstyle=\color{purple},
    numberstyle=\tiny\color{codegray},
    stringstyle=\color{codepurple},
    basicstyle=\ttfamily\footnotesize,
    breakatwhitespace=false,         
    breaklines=true,                 
    captionpos=b,                    
    keepspaces=true,                 
    numbers=left,                    
    numbersep=5pt,                  
    showspaces=false,                
    showstringspaces=false,
    showtabs=false,                  
    tabsize=4
}

\title{Nonparametric causal discovery with applications to cancer bioinformatics} 
\normalsize
\author{\\\vspace{0.25cm}Jean Pierre Gómez Matos} 
\advisor{\large
\\\vspace{0.25cm}Gabriel José Gil Perez \\
\small https://orcid.org/0000-0002-7728-9522 \large
\\\vspace{0.2cm}Augusto de Jesús Gónzalez García \\
\small https://orcid.org/0000-0002-2367-6113 \normalsize}
\degree{Bachelor of Computer Science}
\faculty{Faculty of Mathematics and Computing}
\date{November 2022\\\vspace{0.25cm}\href{https://github.com/jean-pierre-gm/CChains}{github.com/jean-pierre-gm/CChains}}
\logo{Graphics/uhlogo}
\makenomenclature

\begin{document}

\frontmatter
\maketitle

\begin{dedication}
    To each and every one of the people who got me this far. 
\end{dedication}

\begin{acknowledgements}
	I want to thank my parents, who have been by my side unconditionally in each
	and every phase of my life, have silently resisted the problems and difficulties without
	making me a direct participant, and have supported me in every small project, giving
	their all and more to get me ahead. To my grandmother, who sees through my eyes,
	always attentive to everything that happens to me, one of the few who knows how to
	make me smile when I need it most with her witticisms, and who goes out of her way
	for me. On them I place all my affection and love, and all my gratitude. To my tutor
	Gabriel, who has put in a superhuman effort to carry out this project and has become
	the most important person I have met in the last year, and who, more than my
	mentor, I would gladly like to call him my friend. To all my friends, old and new, who
	in one way or another have built the path I have taken to this point: to Daniel Martínez,
	who went from being my friend to becoming the brother I never had, and without
	whom my life would surely not have the same color, to Deby, who has become part
	of me, a confidant of secrets, evils and smiles, in whom I have placed all my trust, as
	she has in me, to Juan Pablo, unconditional friend who has become my arms and
	legs many times, has earned my trust and respect in record time, and to whom I will
	never be able to repay everything he has done for me, to Javier, one of my goals to
	overcome during this race, the person with whom I have unconsciously competed in
	silence, one of those I admire most for his ability, Daniel Rivero, the person who,
	along with Javier, comes to mind when I hear the word genius, with whom it has been
	a pleasure to work throughout this entire period. To each and every one of them,
	thank you very much 
\end{acknowledgements}

\begin{abstract}
	Many natural phenomena are intrinsically causal. The discovery of the cause-effect relationships implicit in these processes can help us to understand and describe them more effectively, which boils down to causal discovery about the data and variables that describe them. However, causal discovery is not an easy task. Current methods for this are extremely complex and costly, and their usefulness is strongly compromised in contexts with large amounts of data or where the nature of the variables involved is unknown. As an alternative, this paper presents an original methodology for causal discovery, built on essential aspects of the main theories of causality, in particular probabilistic causality, with many meeting points with the inferential approach of regularity theories and others. Based on this methodology, a non-parametric algorithm is developed for the discovery of causal relationships between binary variables associated to data sets, and the modeling in graphs of the causal networks they describe. This algorithm is applied to gene expression data sets in normal and cancerous prostate tissues, with the aim of discovering cause-effect relationships between gene dysregulations leading to carcinogenesis. The gene characterizations constructed from the causal relationships discovered are compared with another study based on principal component analysis (PCA) on the same data, with satisfactory results.
	  
\end{abstract}

\tableofcontents
\listoffigures
\listoftables

\mainmatter

\chapter*{Introduction}\label{chapter:introduction}
\addcontentsline{toc}{chapter}{Introduction}
Causality is a fundamental concept in our way of understanding the world.
It appears as a subject of study by Aristotle in his book \textit{Analytical Seconds}, but it was
in the 18th century that it acquired its first detailed formulation from the hand of David
Hume. According to this author, causality is associated with the achievement of two
events, contiguous in time and space.\brackcite{Hume} Since then, the existence of a necessary
connection between the two has been presumed, which must be explained. It is the
current impossibility of specifying the nature of this necessary connection in empirical
terms that makes the study of causality a true conceptual challenge.

Human intuition seems to be able to determine what is cause and what is effect in a
large number of cases. For example, if a billiard ball hits another, it can be inferred that
the motion of the first is the cause of the subsequent motion of the second. Modern
science has been able to expand this intuition to regions and scales of the universe
inaccessible to our perception, revealing the causal mechanisms behind certain
phenomena. To mention one case, today it is known that certain proteins are responsible
for the translation of the genetic code into others, even when this is not a process
observable to the naked eye. However, there are cases where not even the support of
modern science is enough to establish causal relationships. For example, the causal
relationships established between socioeconomic indicators are not always easy to
explain in a mechanistic framework, and therefore it is not easy to predict the influence
of specific policies on them.

This scenario has guaranteed that in recent years discovery and causal inference
have gained scientific recognition. In fact, the 2021 Nobel Prize in Economics was
awarded to J. Angrist and G. Imbens for their contributions to the methodology for the
analysis of causal relationships.\brackcite{Nobel} Also, the field of causal modeling has been
developed, which brings together a set of techniques to re-present systems of causal
relationships and infer them from probabilities. One of the most important figures within this field is Judea Pearl, 
who created the computational basis for information processing under uncertainty. He is also credited
with the invention of Bayesian networks, as well as the main algorithms used for
inference in these models. Their work revolutionized the field of artificial intelligence,
and overturned the long-held belief that causation can only be determined from randomized controlled trials, 
which isimpossible in areas such as biological and social sciences.\brackcite{Turing} His contributions to
causal modeling earned him the 2011 Turing Award, \textit{"for his fundamental contributions
to artificial intelligence through the development of a calculus for probabilistic and
causal reasoning"}.\brackcite{Turing2}

Nowadays, there is no doubt about the potential and importance of causal
discovery. However, most algorithms designed for this use complex and extremely
expensive methods. They generally consist of parametric methods, ie, they make
assumptions about the probability distribution in the data, and are not practical
solutions for large volumes of data. This problem lies in a large part of the motivations
and objectives of this study, which proposes to design a non-parametric algorithm for
causal discovery, which allows detecting cause-effect relationships between the
variables associated with a set of data from a different approach than the usual ones,
focusing mainly on \textit{sufficient causal relationships}. As a secondary motivation, we want
to apply this algorithm to discover causal relationships between genetic alterations
associated with carcinogenesis. The final objective is to discover genes whose
alteration serves as an indication of the presence of cancer (ie, indicator genes), and
genes whose alteration plays a fundamental role in the regulatory mechanisms that at
the genetic level promote the origin and proliferation of said disease. (ie, target genes).
The detection of genes with these characteristics is of high importance for the
diagnosis and treatment of cancer. In particular, the target genes could be used in
alternative therapies to conventional chemotherapy, since the suppression or
replacement of their genetic expressions in cancerous tissue could contribute to the
regulation of the disease, and its possible cure.

Therefore, the general objective of this work can be summarized:
\begin{itemize}
	\item Proposal of a causal discovery methodology to detect causal relationships
	between variables referring to a set of data, and its application for the discovery
	of cause-effect relationships between genetic alterations that cause and regulate
	carcinogenesis.
\end{itemize}
Therefore, its field of action mainly covers causality, probabilities, graph theory and
genomics. Its specific objectives are:
\begin{itemize}
\item Analyze the main characteristics of causal relationships, and some of the
theories that define and study them.
\item Study current methods of causal discovery, paying special attention to the
modeling of causal relationships using graphs.
\item Analysis of the behavior of causal relationships in cases of causal sufficiency.

\item Design a causal discovery algorithm that:
\subitem $\bullet$ is non-parametric,
\subitem $\bullet$ describes sufficient causal relationships, 
\subitem $\bullet$ does not employ statistical tests of conditional independence,
\subitem $\bullet$ minimize the number of spurious, redundant, and unknown-direction causal
relationships represented in the resulting model.
\item Apply this algorithm in the construction of genetic regulation networks associated
with carcinogenesis, in order to identify cancer indicator genes and possible
target genes for therapy.
\item  Develop software that implements the algorithm and functionalities related to
causal graph analysis, efficiently and accessible to users, with particular attention
to those within the field of bioinformatics.
\end{itemize}  

The chapters of this document are organized as follows:
\begin{itemize}
\item \textbf{Chapter 1}: The different theories that define and study causal relationships are
analyzed. An approach is made to current methods of causal discovery, paying
particular attention to the modeling of causal relationships in sets of variables
through graphs.
\item \textbf{Chapter 2}: A brief analysis of the main characteristics of causal relations,
particularly sufficient causal relations, is carried out.

Chapter 5: The conclusions of the research are presented.
A methodology and algorithm is proposed for the discovery of sufficient causal
relationships in a set of variables associated with a population of individuals.
\item \textbf{Chapter 3}: An implementation of the proposed algorithm is described, and the
main characteristics of the program that includes it.
\item \textbf{Chapter 4}: The results obtained after applying the proposed algorithm on a data
set of genetic alterations associated with cancer, particularly prostate cancer,
are analyzed.
\item \textbf{Chapter 5}: The conclusions of the research are presented.
\end{itemize}

\chapter{State of the art}\label{chapter:state-of-the-art}

Below is an analysis of the main theories that study causality and its properties. These
theories are various and complex, with extensive precedent in various fields such as Physics
and Philosophy. The study is reduced to a subset of them, which will serve as the basis for
the methodology to be developed in the following chapters. 

\section{Regularity theory}\label{regteo}
The theory of causal regularity is based on the observation that causes are regularly
followed by their effects. This theory has a reductionist approach, and does not propose
causal powers by virtue of which causes produce their effects.
The Scottish empiricist philosopher David Hume (1711-1776) is the father of this conception
of causality. According to this author, cause and effect\brackcite{Hume}\\

\textit{``(...) are [objects] contiguous in time and place and that the named object causes precedes the other, which we call effect.''}\\

From his perspective, cause and effect appear in a kind of constant conjunction but there
is no necessary connection between them. In fact, his theory attributes to the mind the idea
that, by necessity, a cause is followed by its effect:\brackcite{Hume}\\

\textit{``(...)  after frequent repetition I see that when one of the objects appears the mind is
	determined by habit to attend to its usual companion, and to consider it (...), by virtue of its
	relationship with the first object. It is then this impression or determination that gives me
	the idea of necessity.''}\\

\noindent For Hume, the impression of a necessary connection emerges as a result of the inferential habits developed on the basis of experience of regularities. However, later approaches to regularity theory conceive of causality as an objective process, independent
of an observer's experience.

In summary, Hume's regularity theory states that, let $c$ and $e$ be two instances of events
of types $C$ y $E$, respectively, $c$ is a cause of $e$ if and only if\brackcite{Reg}:

\begin{itemize}
	\item $c$ is spatially and temporally contiguous to $e$,
	\item $c$ precedes $e$ in time, and
	\item all type $C$ events are followed by an event of type $E$. 
\end{itemize}

\noindent The first condition excludes causal relations at a distance (except, perhaps, through a
causal chain of events), the second establishes that the cause-effect succession is oriented
in the same direction of time, and the third indicates the manifestation of cause and effect
in constant conjunction\brackcite{Reg}. 

This theory has some advantages. On the one hand, it does not require an objective
but inaccessible explanation that can support the idea of a necessary connection between
cause and effect; on the other, it reduces causality to three verifiable conditions
(spatiotemporal contiguity, temporal precedence, and regular association). In this way,
Hume's theory allows causal relationships to be inferred in a relatively simple way. Since
temporal precedence and spatiotemporal contiguity are generally observable, identifying
causal relationships is reduced to the mere detection of regularities\brackcite{Reg}.

However, this approach to causality presents two fundamental problems:
\begin{itemize}
	\item It does not contemplate \textit{singular} causal relationships.
	\item Introduces \textit{spurious} causal relationships.
\end{itemize}

By singular causal relationships, we understand any causal relationship that takes
place only once. For example, the death of Archduke Franz Ferdinand (1863 - 1914), heir
to the Austro-Hungarian crown, is recognized as the cause of the First World War (although
not the only one)\brackcite{War}. These events happened only once in History and are, in fact,
unrepeatable (like any other historical event). However, many agree that there is a cause-effect 
relationship between the two, contradicting the notion of causality as an inferential habit when identifying a regularity.

On the other hand, an example of a spurious causal relationship is the association of
the light on while sleeping as a cause of myopia in young children. In 1999, a team from
the University of Pennsylvania (USA) published a study in  \textit{Nature}which concluded that babies who sleep with the light on were at greater risk of developing myopia, indicating a possible cause-effect relationship between these events. 
However, a subsequent study from Ohio State University refuted these claims, finding no such
causal relationship between the two and, instead, revealing a tendency in myopic parents to
leave the lights on in their children's rooms, as well as a hereditary factor in this disease\brackcite{Myopia}. In
this case, a regularity occurs, and yet there is no causal relationship between infants' myopia
and the light on when sleeping. In fact, the correlation between both events is due to a common
cause: paternal myopia.

The existence of non-causal regularities shows that the definition of causality proposed by
Hume is, to say the least, incomplete. Addressing this problem, Mill (1806 – 1873) refined
Hume's theory of causality. Mill extends the concept of cause to a conjunction of positive and
negative factors, where the presence of the positive and the absence of the negative produces
the appearance of the effect. On the other hand, he states that mere regularity is not sufficient
to determine causality. The regularities of causality require support from some law of nature\brackcite{Reg}. 
By law of nature, Mill understands a set of more general regularities\brackcite{Mill}:\\

\textit{``[The term] Laws of Nature means nothing more than the uniformities that exist among
	natural phenomena (or, in other words, the result of induction) when they are reduced to their
	simplest expression.''}\\

\noindent Note that in his definition of laws of nature there is an implicit human factor. 
Appreciates regularities as natural processes, but their status as \textit{law} depends on subjective
criteria of generality and simplicity.

Mackie (1917 – 1981) is one of the heirs of the regularist tradition in the 20th century.
According to its formulation, each of the groups of factors of a cause is a conjunction of positive
terms (eg, the presence of events) and negative terms (eg, the absence of events). Then, the
disjunction of all these conjunctions re-presents the totality of causes of an event. For example,
a lit match, the presence of dry leaves and the absence of rain can be the causes of a forest fire,
as can the breakdown of a power line and a delay in the deployment of preventive actions, or an
accident. automobile and the presence of a fuel shipment in one of the vehicles involved, etc.
Mackie proposes that each of the factors of a conjunction is a cause, but\brackcite{Mackie}\\

\textit{``[t]he so-called cause is, and is known to be, an insufficient but necessary part of a condition
	which is itself unnecessary but sufficient for the result. In view of the importance of conditions of
	this kind in our knowledge and ways of talking about causality, it will be convenient to have a
	short name for them: INUS conditions.''}\\ 

\noindent Mackie proposes that a cause $C$ is always an INUS (Insufficient-Necessary-Unnecessary-Sufficient) condition of the associated effect $E$, admitting the exceptions where $C$ is by itself a
sufficient cause, a necessary cause, or both. Therefore, from their perspective, a cause can
have essentially four forms. Let  $C_{i,j}$ be a causal event on an effect $E$, and '/$\rightarrow$' be the symbol of
the causal relationship: 

\begin{itemize} 
	\item si $C_{1,1} \rightarrow E$, then $C_{1,1}$ is a unique, necessary and sufficient cause;
	\item si $(C_{1,1} \wedge C_{1,2} \wedge ...C_{1,n}) \rightarrow E$, then $C_{1,1}$ is a necessary cause, but not sufficient;
	\item si $C_{1,1} \vee (C_{2,1} \wedge C_{2,2} \wedge ...C_{2,m}) \vee ...(C_{k,1} \wedge C_{k,2} \wedge ...C_{k,l}) \rightarrow E$, then $C_{1,1}$ is cause sufficient, but not necessary;
	\item si $(C_{1,1} \wedge C_{1,2} \wedge ...C_{1,n}) \vee (C_{2,1} \wedge C_{2,2} \wedge ...C_{2,m}) \vee ...(C_{k,1} \wedge C_{k,2} \wedge ...C_{k,l}) \rightarrow E$, then $C_{1,1}$ is an INUS condition.
\end{itemize}

\noindent Mackie's theory, although it presents great facilities for causal inference, and denotes causes
more explicitly and concisely than its predecessor Mill, is incapable of distinguishing spurious
causal relationships. On the other hand, he raised doubts about an apparent symmetry of
complex regularities from his approach, which contradicts the asymmetry inherent to causal
relationships in general\brackcite{Reg}.    

Baumgartner subsequently denies claims about such symmetry. From his point of view,
Mackie's complex regularities remain asymmetric, in the sense that the instantiation of any
conjunction $(C_{i,1} \wedge ...C_{i,n})$ of causes determines the effect $E$, as instantiating $E$ it does not
allow us to determine more than the total disjunction of all these conjunctions. Baumgartner
defends what is known as the asymmetry of overdetermination, in which any of the sufficient
causes of an effect determine it, but the effect cannot determine any of its sufficient causes
separately. This author states that the main deficiency in Mackie's theory lies in not strictly
minimizing material regularities, and states that\brackcite{Baumgartner}\\

\textit{``[t]he most important condition that regularities must satisfy to be causally interpretable is
	what can be called a principle of non-redundancy. The causal structures do not present
	redundancies. Each cause contained in a causal structure [...] makes a difference in at least
	one effect of the structure, in at least one situation.''}\\

\noindent According to Baumgartner, to characterize the causal structure of a complex regularity, a
minimally necessary disjunction of minimally sufficient conjunctions is required.

Sometimes, theories of causal regularity have become inferential theories of causality. Inferential theories propose that a causal relationship can be taken as refinements and generalizations of regularity theory. For example, an INUS condition can be explained in terms of logical inferences, interpreting the causal relation '$\rightarrow$' as a biconditional '$\iff$'\brackcite{Reg}:\\

\begin{center}
	$(C_{1,1} \wedge C_{1,2} \wedge ...C_{1,n}) \vee (C_{2,1} \wedge C_{2,2} \wedge ...C_{2,m}) \vee ...(C_{k,1} \wedge C_{k,2} \wedge ...C_{k,l}) \iff E$ ~.
\end{center}

\section{Probabilistic causality theory}
\label{ProbCausation}
The theory of causal regularity generally presents a deterministic perspective, assuming
that causes occur together with their effects in invariable succession. However, a cause-effect relationship may exist between two events without necessarily always appearing together. For example, Covid-19 causes loss of smell, difficulty breathing, lung damage, among others, but there are many cases of asymptomatic patients. This imperfection in the infection-symptoms regularity can be explained within the framework of the theory of causal regularity, ie, the infection could be an INUS condition of the symptoms, and the apparent imperfection would then arise from not considering the rest of the causes (positive and negative) 
that must be presented in conjunction with this (genetic factors, immune response and comorbidities of the infected person, etc.). Although this deterministic approach can be useful in some contexts, usually the number of causal factors of an event it's huge. Hence the need to limit the analysis to a few relevant factors, inevitably losing part of the causal information. All phenomenology described only by these relevant factors behaves stochastically. This motivates the proposal of a theory that evaluates causality through probabilities. 

The probabilistic theory of causality is based on the premise that causes increase the probability of their effects. Let $C$ be the cause and $E$ be the effect, then it is required that:  

\begin{equation}
	P(E\mid C) > P(E) ~,\label{pc1}
\end{equation}

\noindent that is, the probability of the effect is greater given the cause. In another formulation is

\begin{equation}
	P(E\mid C) > P(E\mid \neg C) ~,\label{pc2}
\end{equation}

\noindent so the increase in the probability of the effect given the cause is measured with regarding the condition in which the cause is absent. It can be shown that the two inequalities (\ref{pc1})-(\ref{pc2}) are equivalent\brackcite{Cartwright}. 

This basic definition has certain advantages. First, one event can increase the probability
of another, and therefore be its cause, without the need for a constant conjunction (see previous section). 
Therefore, it admits imperfections in regularities (or irregularities) naturally. On the other hand, if $C$ is the cause of $E$ then $C$ is in some way relevant to $E$ (its probability increases), condition not necessarily present in the regularities in general. However, the probabilistic theory based
only on (\ref{pc1})-(\ref{pc2}) does not solve the problem of spurious causal relationships: 
it can be true that $P(E\mid C) > P(E \mid \neg C)$, even when $C$ does not cause $E$; for example, if both are effects of a common cause. Nor does it reflect the asymmetry of causality, since $P(E\mid C)>P(E\mid \neg C)$ if and only if $P(C\mid E)>P(C\mid \neg E)$. Therefore, conditions (\ref{pc1})-(\ref{pc2}) are not sufficient to determine whether $C$ is a cause $E$ o vice versa.\brackcite{Prob}

Reichenbach (1891 – 1953), father of the theory of probabilistic causality,
addresses the problem of spurious relationships based on a condition that he called \textit{screening off}.
$A$ is said to screen $C$ from $E$ when

\begin{equation}
	P(E \mid  A \wedge C) = P(E\mid A) ~,
\end{equation}

\noindent which is equivalent to $P(C \wedge E \mid  A) = P(C\mid A)P(E\mid A)$ if $P(C \wedge E) > 0$. That is, conditioning in $A$ breaks the correlation between events  $C$ and $E$.\brackcite{Reichenbach}. 
A screening off can happen in two ways:

\begin{itemize}
	\item $C$ causes $A$, $A$ causes $E$,  and there is no other path other than $A$ by which the event $C$ can affect event $E$.  In this case, the causal relationship from $C$ to $E$ is considered \textit{indirect} o \textit{remote}, through the mediator $A$, as opposed to the \textit{direct} or \textit{near} causality that is established between cause and effect when there are no other intervening events.
	
	\noindent For example, exposure to SARS-CoV-2 carriers can cause contagion through contact
	with infected secretions. However, among the people who have contact with infected secretions, exposure or not to SARS-CoV-2 carriers does not influence the probability of infection. In addition, if the healthy person were protected by an effective mask, thus preventing own mucous membranes come into contact with foreign secretions, the probability contagion does not depend on \textit{direct} exposure to SARS-CoV-2 carriers.
	
	\item $A$ is the common cause of events $C$ y $E$, and between $C$ and $E$  there is no other causal connection.
	
	\noindent For example, there is a correlation between the light on when sleeping and the onset 
	of myopia in young children. However, since the cause of both events is usually paternal myopia,
	among children whose parents suffer from myopia, those who sleep with the light on are no
	more likely to develop this disease than those who sleep in the dark.
\end{itemize}

Reichenbach considers that $C$ ss \textit{relevant cause} of a subsequent event $E$ if:

\begin{itemize}
	\item $P(E\mid C) > P(E)$, and
	\item there is no set of events prior to or simultaneous with $C$ that screen $C$ from $E$. 
\end{itemize}

\noindent The second condition excludes the possibility that there is a common cause that explains the
correlation between $C$ and $E$, dictated by the first condition. Note that the second condition refers to
events prior to or simultaneous with $C$. Therefore, the shielding mentioned cannot be caused by an
intermediary cause. In the latter case, $C$ would be an indirect (but not irrelevant) cause of $E$.

Additionally, Reichenbach arrives at a series of conditions, known collectively as the ``principle
of common cause'', which are verified in the event that the correlations between events are not due
to a causal relationship between them. According to this principle, if $E_1$ and $E_2$ are positively
correlated (ie, $P(E_1 \wedge E_2) > P(E_1)P(E_2)$) but do not form a cause and effect pair, then they are effects of a common cause $C$ and are meet the following conditions:\brackcite{Reichenbach}

\begin{enumerate}
	\item $0 < P(C) < 1$	
	\item $P(E_1 \wedge E_2\mid C) = P(E_1\mid C)P(E_2\mid C)$ \label{screening}
	\item $P(E_1 \wedge E_2\mid \neg C) = P(E_1\mid \neg C)P(E_2\mid \neg C)$ \label{screening_not}
	\item $P(E_1\mid  C) > P(E_1 \mid  \neg  C)$ \label{e1_father}
	\item $P(E_2 \mid  C) > P(E_2 \mid  \neg  C)$ \label{e2_father}
\end{enumerate}

\noindent Inequalities (\ref{e1_father})-(\ref{e2_father}) are satisfied by virtue of the fact that $C$ is the common cause of $E_1$ and $E_2$. 
Now, in (\ref{screening})-(\ref{screening_not}) it is stated that $C$ and $\neg C$ screen the relationship between $E_1$ and $E_2$, so that $C$ is responsible for the correlation between $E_1$ and $E_2$.

The definition of cause proposed by Reichenbach, in conjunction with this principle,
solves the problems of spurious causal relationships, while making use of the order
temporal to explain the asymmetry of cause and effect. However, it is not a sufficient
condition of causality. Consider, for example, the causal relationship between
contraceptive pills and thrombosis. Birth control pills can cause thrombosis episodes.
However, these drugs prevent pregnancy, which constitutes an even greater risk of thrombosis.\brackcite{Hausman} This is a controversial example and must not be taken as an empirically demonstrated reality. However, and since is illustrative, in this context it is assumed to be true. Let $C$ be the ingestion of birth control pills, $E$ be the increase in blood pressure and $A$ be pregnancy, then it can be given that $P(E\mid C) < P(E\mid \neg C)$ (and therefore, that $P(E\mid C) < P(E)$). This means that that the ingestion of birth control pills can indirectly prevent the \textit{increase} in blood pressure, by \textit{directly} preventing pregnancy. 

With inequality (\ref{pc1}) reversed, $C$ could not be considered a relevant cause of $E$ according to the definition of Reichenbach, but $C$ is the cause of $E$ (and is not irrelevant). Note that if it is conditioned on $A$ or $\neg A$, then it is possible to obtain the correct inequalities, ie, $P(E\mid C \wedge A) > P(E\mid \neg C \wedge A)$ y $P(E\mid C \wedge \neg A) > P(E\mid \neg C \wedge \neg A)$, in each case. Indeed, contraceptive pills raise the blood pressure of both pregnant and non-pregnant
women. This problem is known as the Simpson's paradox.\brackcite{Cartwright}

In this scenario, Cartwright proposes a new definition of cause, based on conditioning on background contexts. A background context is nothing more than a conjunction of factors (events). This author proposes that $C$ causes $E$ if and only if:

\begin{equation}
	P(E\mid C \wedge B) > P(E \mid  \neg C \wedge B) 
\end{equation}
\noindent for every background context $B$, such that $B$ is formed by different causes of $E$ of $C$ and the effects of $C$.\brackcite{Cartwright} Skyrms, on the other hand, proposes a weaker condition, $C$ causes $E$ if and only if a background context $D$ exists, such that:
\begin{equation} 
	P(E\mid C \wedge D) > P(E \mid  \neg C \wedge D) 
\end{equation}
\noindent and for any background context $B$ it holds $P(E\mid C \wedge B) \geq P(E\mid \neg C \wedge B)$.\brackcite{Skyrms}. On the other hand, Dupré suggests an intermediate condition. It proposes that $C$ should be considered a cause of $E$ if the probability of E increases in a representative sample of the target population, ie,
\begin{equation}  
	\sum_B P(E\mid C \wedge B)P(B) > \sum_B P(E\mid \neg C \wedge B)P(B) ~.
\end{equation}

\section{Theories of manipulability}\label{manteo}
Manipulability theories are based on the idea that manipulating causes must be a way of
manipulating effects. The first approaches to this theory gave a relatively central role to man,
and were reductionist in nature.

One of the prominent theories in this framework was the theory of Menzies and Price,
which attempts to reduce causality using free action as a primitive notion. The definition of
free action is not entirely clear, but it seems to refer to an action without a cause or a
deterministic cause, or to an action that comes from the voluntary choices of an agent.\brackcite{Mani}
On this basis, these authors take an event $C$ to be the cause of another event $E$ if a free agent
can cause the occurrence of $E$ by guaranteeing the occurrence of $C$.\brackcite{Menzies} The objective 
of this link between causality and free agency is to support probabilistic analysis by replacing
traditional conditional probabilities with probabilities conditioned on a free action that
guarantees the occurrence of an event: $C$ is considered the cause of $E$ if the probability of $E$
is greater given that $C$ is obtained through a free act.\brackcite{Mani} This notion of agent has a
fundamentally human focus, and is based on the basic premise that everyone has an
experience of acting as agents, from an early period of life\brackcite{Menzies}. So the analysis of causality in terms of agency preserves the reductionist ideal, and does not suffer from circularities.
However, this makes the interpretation of causality difficult or impossible in those scenarios
where human action is not conceivable. Menzies and Price try to solve this problem by
alluding to the similarity between these non-manipulable processes and others that can be
manipulated, such as event simulations, as long as the latter capture the essential
characteristics of the former. The problem with this definition is that it is implicitly causal, and
therefore not reductive at all.\brackcite{Mani} In general, analyzes of causality through free action 
fail in contexts where agents are not operative. Furthermore, they are incapable of solving the
problem of spurious relationships: the free action that triggers an event $C$ could be correlated
with, or properly be, a common cause of $C$ and $E$, causing a false notion of a cause-effect
relationship between $C$ and $E$.

Before entering into Judea Pearl's interventionist theory\brackcite{Pearl}, it is worth adding a short
digression. There is still unresolved controversy about what causality relates; In other words,
what type of entities belong to what we call cause and effect. Different authors refer to events
(eg, Lewis\brackcite{Lewis}), facts (eg, Mellor\brackcite{Mellor}), conditions (eg, Mackie\brackcite{Mackie}), etc. Until now, it has been useful and intuitive to discuss regularist, probabilistic and manipulation theories in terms of events without, necessarily, their respective authors having shared this choice. Within the framework of Pearl's theory, however, causality relates variables. Either way, when variables are binary, their values can identify the occurrence or not of a certain event.

Pearl brings a new approach to the theory of manipulability, starting from the
concept of intervention. According to this author, an intervention is an atomic alteration
to a variable $X_i$, which does not directly affect any other variable in the system. The
objective of these interventions is, therefore, to measure the effect that a variable $X_i$
has on the rest: any change in another variable cannot be an effect of the intervention
and will, therefore, be a product of the change in the value of $X_i$, revealing a relationship causal.
This approach, unlike its predecessors in manipulability theory, is not reductive, as it
accepts intervention as a causal notion.\brackcite{Mani} 

Pearl models the causal relationships between a set of variables as a directed
graph and an associated system of equations. Each equation of the system represents
an autonomous causal mechanism, and has the form $X_i = F_i(Pa_i, U_i)$, where $Pa_i$ are
the direct causes of $X_i$ (parents of $X_i$ in the graph) that are explicitly represented as
variables of the set, and $U_i$ is an error (or noise) variable that measures the influence
that all the variables not considered have on $X_i$.\brackcite{Pearl} The autonomy of each individual
mechanism is weighed against the possibility of disturbing it (and therefore the
associated equation) without affecting the rest.\brackcite{Mani} Then:\brackcite{Pearl}\\

\textit{``The simplest type of external intervention is one in which a single variable, say $X_i$, is forced to assume some fixed value $x_i$. Such an intervention, which we call ``atomic'', amounts to removing $X_i$ from the influence of the old functional mechanism $X_i = F_i(Pa_i, U_i)$ and placing it under the influence of a new mechanism that sets the value $x_i$ while keeping all other mechanisms unchanged. Formally, this atomic intervention, which we denote by $do(X_i = x_i)$, or $do(x_i)$ for short, amounts to removing the equation $X_i = F_i(Pa_i, U_i)$ from the model and substituting $X_i = x_i$ into the remaining equations. The new model thus created represents the behavior of the system under the intervention $do(X_i = x_i)$ and, when solved for the distribution of $X_j$, yields the causal effect of $X_i$ on $X_j$.''}\\

Therefore, it assumes that any variable other than $X_i$, whose value is affected by
this intervention, is an effect of $X_i$.\brackcite{Mani}

Pearl defines the causal effect of $X_i$ on another variable $X_j$ by the function $P(X_j|hat X_i) = P(X_j | do(X_i = x_i))$, for possible values $x_i$ of $X_i$.\brackcite{Pearl} Note that $P(X_j|do(X_i = x_i)) \neq P(X_j | X_i = x_i)$, i.e., conditioning on the information that $X_i$ was observed to take the value $x_i$ is not the same as conditioning on the information that $X_i$ was manipulated to take the value $x_i$ (something Menzies and Price had already noticed). For example, if between $X_i$ and $X_j$ there is no causal relationship, but they are correlated by a common cause $X_k$, then $X_i$ and $X_j$ are probabilistically dependent and $P(X_j|X_i=x_i) \neq P(X_j)$. Instead, $P(X_j|do(X_i=x_i)) = P(X_j)$, since by establishing the value of $X_i$ through an intervention all relation of $X_i$ with its causes is broken, in particular its relation with $X_k$ and, therefore, with $X_j$ is broken ($X_i$ and $X_j$ are independent after the intervention).\brackcite{Mani}  

A difficulty with this definition of causal effect is precisely that intervening to change the value of a variable $X_i$ keeps the rest of the equations unchanged. In particular, all causes of $X_j$ other than $X_i$ and effects of $X_i$ remain unchanged. Therefore, when considering the causal effect of $X_i$ on $X_j$ one does not only evaluate the effect that $do(X_i=x_i)$ has on $X_j$, but the contribution between it and the rest of these causes.Ê Prompted by this problem, among others, Woodward and Hitchcock decide to take a different approach to the intervention proposed by Pearl. From their perspective, an intervention of one variable $X_i$ must be defined with respect to another variable, with the further idea of characterizing the causal effect of one variable $X_i$ on another $X_j$ by intervening $X_i$ with respect to $X_j$.\brackcite{Mani} 

\section{Graphs}
In the previous section it was mentioned the use of graphs for the representation of causal relationships. Although the concept of graph is common and recurrent in several fields, an approach to the main definitions surrounding this structure is convenient.

A graph $G$ is a pair of sets $<V, E>$, where the elements of $V$ are called vertices (or nodes), and the elements of $E$ are pairs $<v, w>$ of different elements of $V$, which are called edges (or links). Two vertices are said to be \textit{adjacent} if they are connected by an edge. An edge $<v, w>$ can be directed (in this case it is called an \textit{arrow} and is denoted by $v \rightarrow w$), or \textit{bidirectional} (also called an undirected edge, or just an edge, and denoted by $v \leftrightarrow w$). An arc has an implicit notion of direction, so that the arc $v \rightarrow w$ is not equal to the arc $w \rightarrow v$. On the other hand, the edge $v \leftrightarrow w$ is identical to the edge $w \leftrightarrow v$. 

A graph containing arcs and edges is called a \textit{mixed graph}. A \textit{path} in a mixed graph is a set $P = \{v_1, v_2, ...v_n\}$ of distinct vertices such that $\forall i, 0 < i < n$ exists the arc $v_i \rightarrow v_{i+1}$, the $v_i \leftarrow v_{i+1}$, or the edge $v_i \leftrightarrow v_{i+1}$. If the path $P$ satisfies that $forall i, 0 < i < n$ exists on the arc $v_i \rightarrow v_{i+1}$, then $P$ is said to be a \textit{directed path}. A directed path is called a \textit{cycle} if the arc $v_n \rightarrow v_1$ also exists. 

If a mixed graph contains only arcs then it is a directed graph, while a mixed graph containing only bidirectional edges is an undirected graph. Of particular interest among directed graphs are directed and acyclic graphs ($DAG$). 

Finally, a genealogical relationship is usually established between the vertices of a directed graph. A vertex $v$ is said to be the parent of another $w$ if and only if the arc $v \rightarrow w$ exists. In these cases $w$ is also said to be a child of $v$. A vertex without a parent is said to be an orphan. From this definition, we establish the ancestor-descendant relation for any pair of vertices $(i, j) \in V \times V$, so that:
\begin{itemize}
	\item $i$ is an ancestor of  $j$ if $i$ is the father of $j$.
	\item $i$ is an ancestor of  $j$ if $i$ is an ancestor of $j$'s father.
	\item $j$ is a descendant of $i$ if $i$ is an ancestor of $j$.
\end{itemize}

\section{Bayesian networks}
Given a set $V = \{v_1, v_2, ...v_n\}$ of variables, and a probability function $P$, a Bayesian network $G$ over $V$ consists of a $DAG$ that satisfies: 
\begin{equation}
	P(v_1, v_2, ...v_n) = \prod_{v_i \in V}P(v_i | Pa_i)
\end{equation}
where $Pa_i$ is the set of variables corresponding to the parents of $v_i$ in $G$.

This condition is called \textit{factorization of Bayesian networks}, and in $DAGs$ it is equivalent
to the \textit{local Markov condition}. A graph meets the local Markov condition if every vertex $v_i$
of the graph, conditioned on all its parents, is independent of the rest of the non-descendant
vertices of $v_i$.

The \textit{condition of minimality} is also usually required for A Bayesian network $G$ satisfies the condition of minimality if:\brackcite{Prob}

\begin{itemize}
	\item $G$ satisfies the local Markov condition.
	\item No subgraph of $G$, resulting from eliminating arcs of $G$, satisfies the Markov condition. 
\end{itemize}

Under the Markov condition on a graph G, certain structures within the graph acquire importance. Especially:
\begin{itemize}
	\item A \textit{chain} is a trio of vertices $\{v_1, v_2, v_3\}$ such that the arcs $v_1 \rightarrow v_2$ and $v_2 \rightarrow v_3$ exist.
	\item A \textit{bifurcation} or \textit{fork} is a trio of vertices $\{v_1, v_2, v_3\}$ such that the arcs $v_1 \leftarrow v_2$ and $v_2 \rightarrow v_3$ exist. 
	\item A \textit{immorality} is a trio $\{v_1, v_2, v_3\}$ such that the arcs $v_1 \rightarrow v_2$, $v_2 \leftarrow v_3$ exist, and furthermore the arcs $v_1 \rightarrow v_3$, $v_1 \leftarrow v_3$ do not exist. Within the immorality, the vertex $v_2$ is referred to as the \textit{collider}.
\end{itemize}

\noindent Part of the importance of these structures lies in the relationships of conditional independence
that they fulfill:\brackcite{Neal}
\begin{itemize}
	\item Both a chain and a fork $\{v_1, v_2, v_3\}$ satisfy that $P(v_1, v_3 | v_2) = P(v_1|v_2)P(v_3|v_2)$, that is, $v_1$ and $v_3$  are independent given $v_2$.
	\item An immorality fulfills that:
	\subitem $\bullet$ $P(v_1, v_3) = P(v_1)P(v_3)$, that is, $v_1$ y $v_3$ are statistically independent.
	\subitem $\bullet$ $P(v_1, v_3 | v_2) \neq P(v_1|v_2)P(v_3|v_2)$, that is, $v_1$ y $v_3$ are dependents given $v_2$.
	\subitem $\bullet$ Let $v_4$ be any descendant of $v_2$, then $P(v_1, v_3 | v_4) \neq P(v_1|v_4)P(v_3|v_4)$. 
\end{itemize}

\section{Causal graphs}
Given a set $V = \{v_1, v_2, ...v_n\}$ of variables, a causal graph over $V$ is a graph where each
vertex represents a variable $v_i$ of the set. An edge in a causal graph can be either an arc or a
bidirectional edge. An arc $v \rightarrow w$ represents the causal relationship where $v$ is the cause 
and $w$ the effect, while an edge $v \leftrightarrow w$ denotes the existence of causes common to 
$v$ and $w$ not observed in $V$ (sometimes called confounding factors).\brackcite{Pearl} Therefore, 
if $v$ is the parent of $w$ then $v$ is the cause of $w$.

Usually, the cause-effect relationships represented by arcs in a causal graph are direct
causal relationships, that is, (as defined above in terms of events), those causal relationships
between two variables where there are no other intermediate variables. In these cases, a
directed path represents a chain of direct causal relationships between the variables that
compose it. Therefore, between any two non-consecutive vertices $v$, $w$ of a directed path, the
subpath from $v$ to $w$ represents an indirect causal relationship from $v$ to $w$

\section{PC algorithm for causal discovery}
\label{PC}
A causal discovery algorithm is one that, given a set $V$ of variables, detects the causal
relationships between its elements and represents them consistently, usually, through a causal
graph. As sections \ref{regteo}, \ref{ProbCausation} and \ref{manteo} suggest, the identification of these relationships involves several difficulties. One of the most popular approaches within causal discovery is based on statistical independence as a primitive notion, given that the latter guarantees causal
independence if no other relevant factors exist. In the most general case, conditional
independence is used as a sufficient condition for the absence of a direct causal relationship,
which are the relationships that are usually represented in causal graphs.

Given a graph $G$, associated with a set V of variables and a probability function $P$,
Pearl uses the concept of d-separation to identify relationships of conditional
independence:\brackcite{Pearl}\\

\textit{``A path $p$ is said to be d-separated (or blocked) by a set of nodes $Z$ if and only if:}
\begin{enumerate}
	\item \textit{$p$ contains a chain or a fork such that the middle node $m$ is in $Z$, or}
	\item \textit{$p$ contains an inverted branch (or collider) such that the middle node $m$ 
		is not in $Z$ and such that no descendant of $m$ is in $Z$.}
\end{enumerate}

\textit{A set $Z$ is said to separate [a set] $X$ from [another set] $Y$ if and only
	 if $Z$ blocks all paths from a node in $X$ to a node in $Y$.''}\\

This concept makes sense on the fulfillment of the local Markov condition in the
graph $G$, and has the objective of exploiting the properties of the chains, bifurcations,
and immoralities described above. Pearl states that, for a $DAG$ that meets this condition
(Bayesian network)\brackcite{Pearl}\\

\textit{``If the sets $X$ and $Y$ are d-separated by $Z$ in a $DAG$ $G$, then $X$ is independent of $Y$ conditional on $Z$ in every $G$-compatible distribution. Conversely, if $X$ and $Y$ are not d-separable by $Z$ in a $DAG$ $G$, then $X$ and $Y$ are conditionally dependent on $Z$ in at least one $G$-compatible distribution.''}\\

\noindent This condition is known as the \textit{global Markov condition}, and implies that, given the
structure of $G$, it is possible to discover the conditional independence relations between
the variables of $V$. However, what is desired is an opposite condition, which allows the
structure of the graph $G$ to be discovered from these relations of conditional
independence. This condition is called the \textit{fidelity condition} and dictates that these
independence relations are also necessary for the global Markov condition.
On this basis, Spirtes, Glymour, and Scheines build the $SGS$ algorithm for causal
discovery. Starting from a set $V$ of variables, the $SGS$ algorithm constructs the causal
graph $G$ of $V$ assuming: 

\begin{itemize}
	\item The set of variables $V$ presents causal sufficiency (there are no confounding factors).
	\item The graph $G$ that describes the causal relationships of $V$ is acyclic.
	\item $G$ satisfies the Markov conditions.
	\item The probability distribution of $P$ in the set $V$ is such that the fidelity condition is satisfied.
\end{itemize}

\noindent The algorithm can be summarized in the following steps:\brackcite{Spirtes}
\begin{itemize}
	\item A graph $H$ is constructed on the set $V$ of variables, such that $H$ is an undirected graph
	and contains all possible edges between vertices of $V$. 

	\item For each pair of vertices $v$ and $w$, if there exists a subset $S$ of $V\setminus\{v,w\}$ such that $v$ and $w$ are d-separated given $S$, then the edge between $v$ and $w$ is removed from $H$. 
	
	\item Let $K$ be the undirected graph resulting from the previous step. For every trio of vertices $v$, $w$, $x$ such that the pair $v$, $w$ and the pair $w$, $x$ are each adjacent in $K$ (i.e., $v \leftrightarrow w \leftrightarrow x$), but the pair $v$ and $x$ are not adjacent in $K$, one orients $v \leftrightarrow w \leftrightarrow x$ as $v \rightarrow w \leftarrow x$ if and only if $w$ does not belong to any $S$-subset of $V$ that d-separates $v$ and $x$. 
	
	\item Finally, it is repeated: 
	
	\subitem $\bullet$ for each trio of vertices $v$, $w$, $x$, if $v \rightarrow w$, $w$ and $x$ are adjacent via an undirected edge, and $v$ and $x$ are not adjacent, then $w \leftrightarrow x$ is oriented as $w \rightarrow x$; 
	\subitem $\bullet$ for each trio of vertices $v$, $w$, if there is a direct path from $v$ to $w$, and an undirected edge between $v$ and $w$, then $v \leftrightarrow w$ is oriented as $v \rightarrow w$; 
	
	until no more edges can be oriented.
\end{itemize}

Note that the second step of the algorithm is justified in that, if by conditioning on a set $S$
the dependency between two variables $v$ and $w$ is broken , then between $v$ and $w$ there is no
direct causal relationship and the edge that unites them can be eliminated. 

On the other hand, the third step is based on two essential ideas:\brackcite{Neal}
\begin{itemize}
	\item There exists at least one set $S$ such that $v$ and $x$ are d-separable given $S$.
	
	This statement is true, since at the beginning $G$ contains all possible edges, and the
	edge $v \leftrightarrow w$ is no longer in $G$, then the edge was eliminated by conditioning on some set
	of variables.
	
	\item $w$ does not belong to any set that d-separates $v$ and $x$.
	
	From which it follows that, although $v$ and $x$ are d-separable when conditioning on some
	set, as long as $w$ belongs to the set being conditioned, said d-separation will not be
	achieved. Then $w$ is necessarily a collider, and the subgraph underlying $v \leftrightarrow w \leftrightarrow x$ is $v \rightarrow w \leftarrow x$.
\end{itemize}

This algorithm, although it has a high reliability in terms of causal discovery, presents a high time complexity. The second step requires conditioning, for each pair of vertices connected by an edge, on all possible subsets of remaining variables, thus becoming an exponential time complexity algorithm. In addition:\brackcite{Spirtes}\\

\textit{``In the worst case such complexity is unavoidable if reliability is to be maintained. Two variables may be dependent conditioning on a set $U$ but independent conditioning on a superset or subset of $U$. Any worst-case procedure that does not examine the conditional independence relationships of variables $X$, $Y$ on all subsets of vertices that do not contain that pair will fail.''}\\

\noindent Spirtes, Glymour and Scheines then redesigned the $SGS$ algorithm and the $PC$ algorithm emerged, which essentially replaces the second and third steps of the $SGS$ algorithm with two equivalent, but computationally less expensive, procedures.

\chapter{Proposal}\label{chapter:proposal}
In this chapter, a non-parametric method is proposed for the discovery of causal
relationships between binary variables. The approach used is related to that of the
probabilistic theory of causality, although it is based on different assumptions.

Next, some basic notions of causal relationships are introduced, which will be used in the rest of the chapter

\section{Causal relationships}

Causality is conceived as a binary relationship between variables (cause and effect)
and whose main characteristics are:\brackcite{Hausman}
\begin{itemize}
\item Irreflexivity: For every variable $i$ it is true that $i$ is not the cause of itself.
\item Asymmetry: For every pair of variables $i$ and $j$ it is true that if $i$ is the cause of $j$ then
$j$ is not the cause of $i$. 
\item Transitivity: For every trio of variables $i$, $j$ and $k$ such that $k$ is the cause of $j$ and $j$ is
the cause of $k$, it is true that $i$ is the cause of $k$. 
\end{itemize}

A \textit{causal chain} is defined as a finite sequence of variables $\{i_1, i_2, ...i_n\}, n \geq 2$ such
that $\forall a, 0 \leq a < n$ is true that $i_{a}$ is the cause of $i_{a+1}$.

\subsection{Sufficient, necessary, and contributory causes}

A sufficient cause of a variable $j$ is defined as its cause $i$, such that if $i$ is present then $j$ is also present. Analogously, a necessary cause of a variable $j$ is defined as its cause $i$, such that if $i$ does not occur then $j$ does not occur either. In the first case, the causal relation between $i$ and $j$ is called a sufficient causal relation, and in the second, a necessary causal relation. Contributory causes of $j$ are causes that individually are not sufficient causes of $j$, but whose conjunction with others does constitute a sufficient cause (see state of the art). 

\subsection{Causation by omission and prevention}

Furthermore, the following cases are distinguished: when the cause is identified with the
presence or absence of an attribute or event, and when the effect is identified with the
presence or absence of an attribute or event. The case in which the absence of an attribute
or event produces the effect is known as omission causation. Now, regardless of causality by
omission, if the effect corresponds to the absence of an attribute or event, it is said to be a
case of prevention. Our model ignores both omission causation and prevention. Thus, the
causality implicit in our model connects only the presence of one attribute or event to the
presence of another attribute or event.

\section{Data presentation} 

The starting point is a matrix $M$ of binary variables by individuals, corresponding to the individuals of a population $I$ and variables of a set $V$. Each variable represents the presence or absence of a certain attribute in an individual. In $M$, if $m_{i\gamma}=1$ if the attribute corresponding to the variable $i$ is present in the individual $\gamma$. Likewise, if $m_{i\gamma}=0$ if the attribute that corresponds to variable $i$ is present in individual $\gamma$. Attributes and individuals in the following analysis can be understood in the most general way possible. In particular, individuals can be situations and attributes, events that take place or not in such situations.
	
We define the negation of a variable $i$, and denote it $\neg i$, as a variable that takes value $0$ whenever $i$ is $1$ and takes value $1$ whenever $i$ is $0$.

We denote the frequency of variable $i$ as the frequency with which $i$ takes value $1$ in the population of individuals $I$, i.e., $\pi_i = \frac{1}{\mid I\mid }\sum_{\gamma \in I}m_{i \gamma}$. Similarly, we call the frequency of coincidences of variables $i$ and $j$ the frequency with which $i$ and $j$ take value $1$ in the same individual in the population, i.e., $\pi_{ij} = \frac{1}{\mid I\mid }\sum_{\gamma \in I}m_{i \gamma} m_{j \gamma}$.

In subsequent analyses, we will implicitly use a frequentist interpretation of probability, so that, e.g., $\pi_i$ can be interpreted as the probability that the variable $i$ takes $1$, and $\pi_{ij}$ the probability that $i$ and $j$ take $1$ at the same time. These notions are extended to conditional probability, so that $\pi_{ij\mid \neg k}$ can be interpreted as the probability that $i$ and $j$ take value $1$, given that $k$ takes value $0$. 

\section{Model approach} 
We seek to construct a directed graph $G = <V, E>$, where each vertex corresponds to a variable $i \in V$ and each edge $<i, j> \in E$ represents a causal relationship between pairs of variables $i$ and $j$ of $V$, i.e., a causal graph. Since $G$ is directed, a genealogical relationship is established between its vertices.   

According to the characteristics of causality, $G$ possesses certain properties: 
\begin{itemize}
	\item $\forall i \in V$ is true that $i \rightarrow i \notin E$ (in $G$ there are no ties), due to irreflexivity.
	\item In $G$ there is no more than one arc between two pairs of vertices. If there is direct causality from $i$ to $j$ in different ways, the arc $i \rightarrow j$ represents each and every one.
	\item $\forall (i, j) \in V \times V$, if $i \rightarrow j \in E$ then it must hold that $j \rightarrow i \notin E$, due to asymmetry of causality.
	\item $G$ must be acyclic: any vertex of a cycle, being an ancestor of itself, would be its own cause
	by transitivity, contradicting causal irreflexivity
\end{itemize}

Therefore, $G$ is a simple graph, and must be a $DAG$.

\subsection{Triangles}

For convenience, we also define \textit{triangle} as a triples of vertices $(i, j, k)$, such that there exist the arcs $i \rightarrow j$, $j\rightarrow k$, $i\rightarrow k$. Within a triangle, these arcs are called \textit{lados}. In particular, $i \rightarrow k$ is called the hypotenuse, $i \rightarrow j$ is called the first leg, and $j \rightarrow k$ is called the second leg. Note that the order of the vertices is important, so that the triangle $(i, j, k)$ is not equal to $(i, k, j)$ nor to $(j, i, k)$, for example. 

The measure of length of the edges of a triangle is given by the difference in frequency of its vertices. That is, the length of the arc $i \rightarrow j$ is determined by $|\pi_i-\pi_j|$. Then, the Euclidean classification of triangles according to the length of their edges (scalene, isosceles and equilateral) also applies in this case. Bizarre, but not inconsistent, results arise from this measure of length. For example, it follows that the vertices of an equilateral triangle have equal frequency and that, therefore, their edges have zero length.

A measure for the magnitude of an angle is not available and in the present context is meaningless. In particular, there are no right triangles and the nomenclature of legs and hypotenuse only alludes to the length of the edges and the way a triangle is oriented (whether $(i,j,k)$ or another permutation of the vertex triplet). The length of the hypotenuse must be greater than or equal to the length of the legs. In addition, the hypotenuse shares the parent vertex with the first leg and the child vertex with the second leg. 

\section{Modeling}

Inferring causal relationships is a difficult task. If $i$ causes $j$, then $i$ and $j$ are dependent variables. However, statistical dependence between variables does not ensure that a causal relationship exists between them. In fact, two variables may be statistically dependent on each other due to a common cause. Therefore, it is often said that ``correlation does not \textit{imply} causation''. However, statistical independence between variables is a necessary and sufficient condition for the absence of causal relationships. This is the main motivation of current causal discovery algorithms to employ statistical independence as a primitive notion, and to indirectly infer causal relationships by exclusion rules (see epigraph \ref{PC}). However, in the context of causal sufficiency relations, it is possible to follow another strategy; namely, it is possible to infer causal relations between two variables directly, without resorting to statistical dependence or independence between other variables.

\subsection{Causal sufficiency}
Let $i$ be the cause of $j$. In the deterministic case, the causal relation $i \rightarrow j$ is said to be sufficient if and only if it suffices for the cause $i$ to be present for the effect $j$ to be present. That is, given $i \rightarrow j$, $j$ can be inferred from $i$. So the causal relation $i \rightarrow j$ in causal inference serves the same function as the material implication $i \implies j$ in the \textit{modus ponens} of logic. In the stochastic case, causal inference is not perfect due to the presence of errors, resulting from noise in the data. 

To explain what is meant by errors, examine the truth table corresponding to the material implication:

\begin{table}[H]
	
	\centering	
	\resizebox{200pt}{70pt}{%
		
		\begin{tabular}{ccccc}
			\toprule
			$i$ &		& $j$ &		& $i \implies j$ \\
			\midrule
			\midrule
			0 &		& 0 &	 & 1 \\
			0 &		& 1 &	 & 1 \\
			1 &		& 1 &	 & 1 \\
			1 &		& 0 &	 & 0 \\
			\bottomrule	
		\end{tabular}%
	}	
	\caption{Truth table of the material implication.}
\end{table}

As usual, 0 means false and 1 means true.

The only condition in which the material implication is not fulfilled is in the last row. Therefore, the ordered pair of truth values $(1,0)$ corresponding to the pair of variables $(i, j)$ is called the error (of the material implication). In the error-free case, it is strictly satisfied $i \implies j$. 

If the value space of the variables contains no errors, the variables are logically (and therefore statistically) dependent on each other. When the number of errors is comparable to the number of hits (ordered pairs other than $(1,0)$), the variables are statistically independent. So one can contrast the errors in the data against the errors under conditions of independence to assess how much the material implication is fulfilled and, therefore, how justified the inference represented by $i \rightarrow j$ is. 

\subsection{Loevinger coefficient}
The frequency of errors in $i \rightarrow j$ is $\pi_i - \pi_{ij}$. Now, under conditions of statistical independence $\pi_{ij}=\pi_i\pi_j$. In case of causality, the number of errors made in the inference from $i$ to $j$ is expected to be smaller than in case of statistical independence between the two. From this last condition, we construct the figure of merit that we will use to measure the degree of validity of the causal inference:
\begin{equation}
	\begin{aligned}
		&\pi_i - \pi_{ij} < \pi_i - \pi_i\pi_j\\
		&\pi_i - \pi_{ij} < \pi_i(1 - \pi_j)\\
		&\frac{\pi_i - \pi_{ij}}{\pi_i(1 - \pi_j)} < 1\\
		&1 - \frac{\pi_i - \pi_{ij}}{\pi_i(1 - \pi_j)} > 0\\
		&\frac{\pi_{ij} - \pi_i\pi_j}{\pi_i(1 - \pi_j)} > 0 
	\end{aligned}
	\label{H}
\end{equation}
The expression $H_{ij}=\frac{\pi_{ij} - \pi_i\pi_j}{\pi_i(1 - \pi_j)}$ is known as Loevinger's coefficient. The higher $H_{ij}$, the lower the number of errors. Its maximum value $H_{ij} = 1$ is obtained when $H_{ij} = \pi_{ij}$, and thus corresponds to the error-free case. Its minimum admissible value, $H_{ij} = 0$, is obtained when $\pi_{ij}=\pi_i\pi_j$, and thus identifies the case of statistical independence. It is possible to have $H_{ij} < 0$, but in these cases $\pi_{ij} - \pi_i\pi_j < 0$, so the variables are negatively correlated. It can be shown that cases of negative correlation can only correspond to cases of causation by omission or preemptions:
\begin{equation}
	\begin{aligned}
		&\pi_{ij}-\pi_i \pi_j < 0\\
		&-\pi_{ij}+\pi_i \pi_j > 0\\
		&\pi_i-\pi_{ij}-\pi_i + \pi_i \pi_j > 0\\
		&(\pi_i-\pi_{ij})-\pi_i(1 - \pi_j) > 0\\
		&\pi_{i\neg j}-\pi_i \pi_{\neg j} >0
	\end{aligned}
\end{equation}

Note that the expression $H_{ij}$ is undefined when $\pi_i = 0$ or $\pi_j = 1$ (cases in which the inequality $\pi_i - \pi_{ij} < \pi_i - \pi_i\pi_j$ is not satisfied). In these cases we take $H_{ij}=0$, by convention. Note that causality cannot be inferred if the cause never occurs, or if the effect always occurs.

In principle, it would suffice that $H_{ij} > 0$ be satisfied to assert the existence of causal sufficiency from $i$ to $j$. In practice, a stronger condition, i.e. $H_{ij} > H_0 > 0$, is imposed to increase the level of confidence in this assertion. $H_0$ then constitutes a threshold for causal sufficiency between variables. This is the only parameter of the model. 

On the other hand, the Loevinger coefficient is inherently asymmetric. Namely, $H_{ij} \neq H_{ji}$ if and only if $\pi_i \neq \pi_j$. In particular, $H_{ij} > H_{ji}$ if and only if $\pi_i < \pi_j$. The following is the proof:

\begin{equation}
	\begin{aligned}
		&\pi_i < \pi_j\\
		&\pi_i - \pi_i\pi_j < \pi_j - \pi_i\pi_j\\
		&\pi_i(1 - \pi_j) < \pi_j(1 - \pi_i)\\
		&\frac{1}{\pi_i(1 - \pi_j)} > \frac{1}{\pi_j(1 - \pi_i)}\\
		&\frac{\pi_{ij} - \pi_i\pi_j}{\pi_i(1 - \pi_j)} > \frac{\pi_{ij} - \pi_i\pi_j}{\pi_j(1 - \pi_i)}\\
		&H_{ij} > H_{ji} 
	\end{aligned}
\end{equation}

If $\pi_i < \pi_j$ then $H_{ij} > H_{ji}$. By inverse procedure, it can also be proven that if $H_{ij} > H_{ji}$ then $\pi_i < \pi_j$. 

If $H_{ij} > H_{ji}$ is satisfied, the inference from $i$ to $j$ from $j$ to $i$ makes more sense. Therefore, it is inferred that $i$ is a sufficient cause of $j$ if and only if $H_{ij} > H_0$ and $H_{ij} > H_{ji}$ (in other words, if $H_{ij} > H_0$ and $\pi_i < \pi_j$). While $H_{ij} > H_0$ indicates the presence of sufficient causality, the condition $\pi_i < \pi_j$ indicates its direction. On the other hand, the condition $\pi_i < \pi_j$ is more supported in sufficiency contexts, where an effect is expected to occur more frequently than its cause: the effect may have different causes, and must appear when either of them occurs.  

On the other hand, when $\pi_i = \pi_j$ asymmetry is not satisfied. In fact, if $\pi_i = \pi_j$ then $H_{ij} = H_{ji}$, and both inferences share the same degree of validity. 

The following is the proof:
\begin{equation}
	\begin{aligned}
		&\pi_i = \pi_j\\
		&(1 - \pi_i) = (1 - \pi_j)\\
		&\pi_j(1 - \pi_i) = \pi_i(1 - \pi_j)\\
		&\frac{1}{\pi_i(1 - \pi_j)} = \frac{1}{\pi_j(1 - \pi_i)}\\
		&\frac{\pi_{ij} - \pi_i\pi_j}{\pi_i(1 - \pi_j)} = \frac{\pi_{ij} - \pi_i\pi_j}{\pi_j(1 - \pi_i)}\\
		&H_{ij} = H_{ji}
	\end{aligned}
\end{equation}

In these cases it is impossible to discern \textit{a priori} the cause of the effect
In summary, for a pair $(i,j)$ of variables such that $\pi_i \leq \pi_j$, there are three cases:
\begin{itemize}
	\item $\pi_i < \pi_j$ y $H_{ij} > H_0$, then $i$ is a sufficient cause of $j$. 

	\item $\pi_i < \pi_j$ y $H_{ij} \leq H_0$, then $i$ is not a sufficient cause of $j$. 

Since $\pi_i < \pi_j \implies H_{ij} > H_{ji}$, it is also true that $H_{ji} < H_0$, so that $j$ is not a sufficient cause of $i$ either.

	\item $\pi_i = \pi_j$, so $H_{ij} = H_{ji}$. 

If $H_{ij} = H_{ji} \leq H_0$ then $i$ and $j$ are not causes of each other. 

Otherwise, as the inferences have the same degree of validity, and it is not possible to
identify the direction of causality.

\end{itemize}

Finally, the Loevinger coefficient is equivalent to the measure of causality probabilistic. According to the previous one, it is said that $i$ is the cause of $j$ if it holds:
\begin{equation}
	\begin{aligned}
		&\pi_{j|i} > \pi_j
	\end{aligned}
\end{equation}
Developing inequality:
\begin{equation}
	\begin{aligned}
		&-\pi_{j | i} < -\pi_j\\
		&1-\pi_{j | i} < 1-\pi_j\\
		&\pi_i(1-\pi_{j | i}) < \pi_i(1-\pi_j)\\
		&\pi_i-\pi_{j | i}\pi_i < \pi_i(1-\pi_j)\\
		&\pi_i-\pi_{ij} < \pi_i-\pi_i\pi_j
	\end{aligned}
\end{equation}
which was the starting point for the definition of $H_{ij}$, in \ref{H}.

$Hij$ is also related to the Pearson correlation coefficient. In fact, the Pearson coefficient for
binary variables is the geometric mean of $H_{ij}$ and $H_{ji}$:
\begin{equation}
	\begin{aligned}
		&\root\of{H_{ij}H_{ji}} = \root\of{\frac{\pi_{ij} - \pi_i\pi_j}{\pi_i(1 - \pi_j)} \frac{\pi_{ij} - \pi_i\pi_j}{\pi_j(1 - \pi_i)}}\\
		&\root\of{H_{ij}H_{ji}} = \root\of{\frac{(\pi_{ij} - \pi_i\pi_j)^2}{\pi_i(1 - \pi_j)\pi_j(1 - \pi_i)}}\\
		&\root\of{H_{ij}H_{ji}} = \frac{\pi_{ij} - \pi_i\pi_j}{\root\of{\pi_i(1 - \pi_i)\pi_j(1 - \pi_j)}}\\
		&\root\of{H_{ij}H_{ji}} = \frac{\text{Cov}(i, j)}{\root\of{\text{Var}(i)\text{Var}(j)}}
	\end{aligned}
\end{equation}
The Pearson coefficient is a symmetric version of the Loevinger coefficient, where the validity
of the inferences from $i$ to $j$ and from $j$ to $i$ matter simultaneously.

\subsection{Construction of the causal graph}
\label{Build}
The graph is built on the sufficient causal relations inferred from $H$, on the matrix $M$.
Therefore, in $G$ there is an arc $i \rightarrow j$ if and only if $\pi_i < \pi_j$ y $H_{ij} > H_0$. In cases where $\pi_i = \pi_j$, the arcs $i \rightarrow j$ and $j \rightarrow i$ are placed , as an undirected edge $i \leftrightarrow j$. The purpose of these edges is to reflect that the real direction of causality is unknown. Due to causal asymmetry, it is known that at least one of the two arcs that compose it is spurious, and consequently it must be eliminated (see Section 3.5).

This graph largely satisfies the proposed objectives. However, it presents \textit{a priori} some essential problems. These can be summarized in:

\begin{itemize}
	\item Spurious arches
	\item Non-transitive paths
	\item Redundant arcs
	\item Remaining undirected edges
\end{itemize}

These problems deserve a more detailed treatment, and are the main motivation for the
simplification of the graph $G$, by eliminating specific arcs. It is hoped that this will resolve the problems raised, and obtain a new graph that represents
only sufficient and direct causal relationships.

\section{Spurious arcs}
\label{Reichenbach}

The coefficient $H_{ij}$ is a criterion for inferring causality between a pair of variables $i$ and $j$. As long as $H_{ij} > H_0$ or $H_{ji} > H_0$ a causal relationship is claimed to exist, and is considered to take place from $i$ to $j$ or vice versa, respectively. However, the arc $i ÷to j$ may represent a spurious causal relationship, in which the vertices involved do not form a cause-and-effect pair. 

It turns out that, when $i$ and $j$ are correlated, any of the following three cases may occur:\brackcite{Hausman}
\begin{itemize}
	\item $i$ is cause of $j$
	\item $j$ is cause of $i$
	\item $i$ and $j$ are effects of a common cause.
\end{itemize}

Therefore, the spurious causal relationship between $i$ and $j$ is then due to a cause $k$, common
to both. 
In these cases, for fixed values of $k$, a variation in the values of $i$ should not cause changes
in $j$, or vice versa. That is, conditioning on the values of the common parent $k$ must break the
apparent relationship of causal sufficiency between $i$ and $j$. In other words:

\begin{equation}
	H_{ij | k} \leq H_0 
	\label{Reichenbach1}
\end{equation}

\begin{equation}		
	H_{ij | \neg k} \leq H_0
	\label{Reichenbach2}
\end{equation}

This criterion is inspired by one of the pillars of the probabilistic theory of causality, i.e. Reichenbach's principle of common cause (see Section 2.2).  

Conditions \ref{Reichenbach1} and \ref{Reichenbach2} are sufficient to justify the absence of causal sufficiency from $i$ to \ref{Reichenbach1} y \ref{Reichenbach2} bastan para justificar la ausencia de suficiencia causal de $i$ a $j$ or vice versa. Given a common parent $k$ that meets the aforementioned criterion, it is possible to affirm that $k$ is responsible for the spurious causal sufficiency relationship between $i$ and $j$. Therefore, for every edge $i \rightarrow j \in E$ (including undirected ones), we look to see if there is a cause $k$ common to $i$ and $j$ that satisfies the proposed conditions, and if this is met, the arc $i \rightarrow j$ of $G$ is completely eliminated.

\subsection{Order of elimination of spurious arcs}
\label{Reichenbach_order}
In principle, the order of elimination of spurious arcs is arbitrary. If conditioning on a common
parent and its absence breaks a causal relationship, it is assumed that the causal relationship is
spurious and that the common parent is responsible for the underlying correlation. Now,
accidentally, due to the incompleteness of the data in terms of variables and individuals, the
ordering of the elimination of arcs can be important. First, it is possible that the set of individuals is not sufficient to distinguish a real common father from a fictitious one. In particular, it is possible that the real parent corresponds to a variable external to the graph but is indistinguishable from a variable internal to the graph, given the sample of individuals. The latter leads to situations in which arcs are eliminated that should not have been eliminated based on vertices of the graph. The consequence of this
elimination does not have a local character. That is, there is a cascade effect that results in the
permanence of spurious arcs that could have been eliminated based on the vertices of the graph.

To exemplify the above, all possible cases in which the late elimination of a spurious arc
generates conflicts will be illustrated. Since every spurious arc $j \rightarrow k$ is the second leg 
of the triangle that it shares with $i$, the common parent of $j$ and $k$ responsible for the spurious
relationship, then it is only necessary to consider three cases. These occur when the second leg
of the triangle in question is the first leg, second leg, or hypotenuse of another triangle. Each of
these will be exemplified below, taking a hypothetical triangle $(i, j, k)$ as the triangle to examine.

\begin{figure}[ht]
	\centering
	\includegraphics[height=6cm]{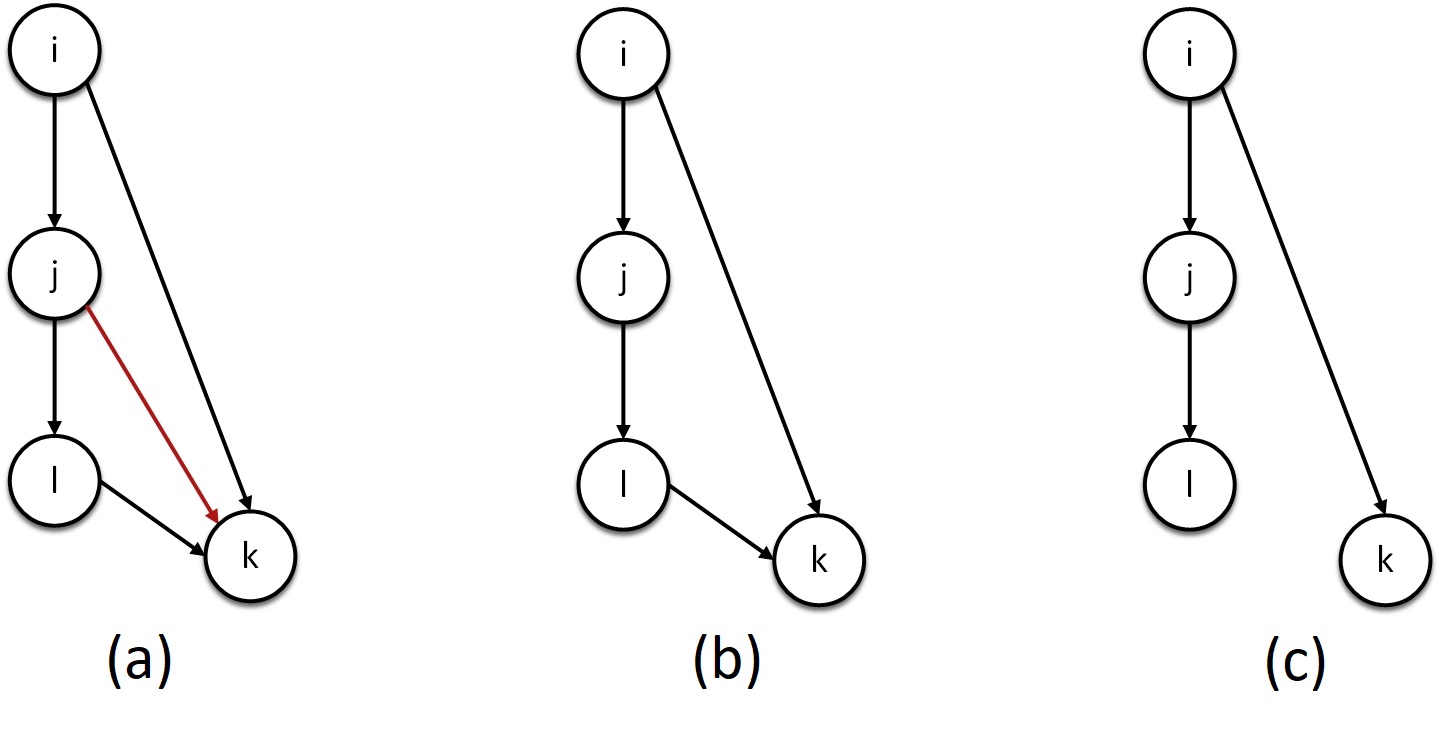}
	\caption{First case of adjacent triangles, with conflict for the elimination of spurious arcs}
	\label{ReichenbachConflict1}
\end{figure}

In the first case, the second leg of triangle $(i, j, k)$ is hypotenuse of another triangle $(j, l, k)$ (\textbf{Fig.}\ref{ReichenbachConflict1}). In this one, both in the diagram \textbf{Fig.}\ref{ReichenbachConflict1}(a), as well as in \textbf{Fig.}\ref{ReichenbachConflict1}(b) and \textbf{Fig.}\ref{ReichenbachConflict1}(c), the frequency of the vertices is decreasing with the ordinate axis (e.g., as $i$ is above $j$ on the $Y$-axis, then $i$ < $j$). Henceforth, this feature will be present in all figures, to gain in explainability and regularity of the diagrams. By construction, in this example the arc $j \rightarrow k$ is spurious and the arc $l \rightarrow k$ must not be eliminated on the basis of graph vertices. That is, the graph underlying \textbf{Fig.}\ref{ReichenbachConflict1}(a) is \textbf{Fig.}\ref{ReichenbachConflict1}(b). Moreover, $i$ is the cause common to $j$ and $k$ that causes the arc $j \rightarrow k$ to appear. If the triangle $(j, l, k)$ is visited first, the arc $l \rightarrow k$ can be eliminated under the erroneous assumption that $j$ is common parent of $l$ and $k$, and then the arc $j \rightarrow k$ by its spuriousness, obtaining the incorrect graph of the \ref{ReichenbachConflict1}(c).

\begin{figure}[ht]
	\centering
	\includegraphics[height=6.5cm]{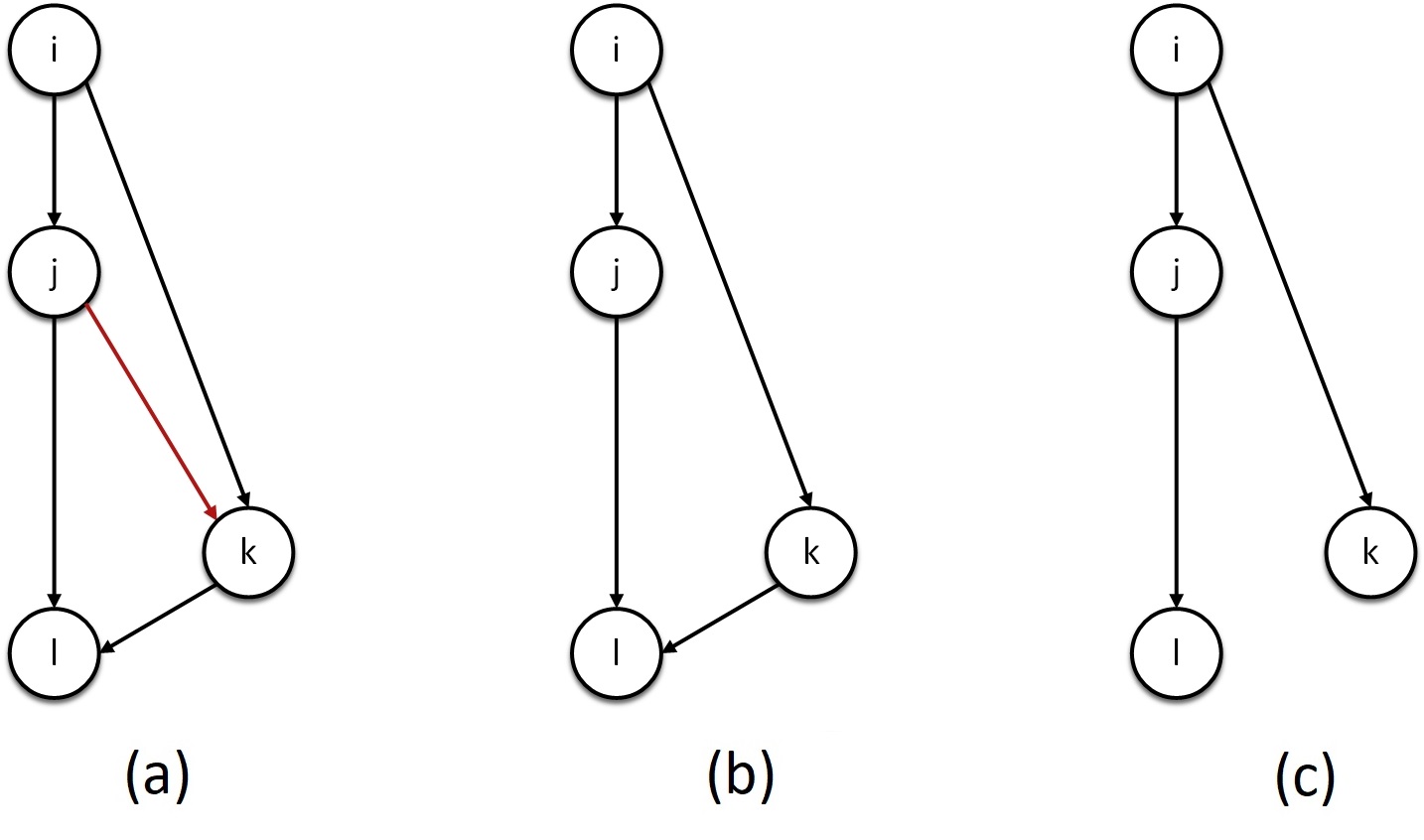}
	\caption{Second case of adjacent triangles, with conflict for the elimination of spurious arcs}
	\label{ReichenbachConflict2}
\end{figure}

In the second case, the second leg of the triangle $(i, j, k)$ is the first leg of another triangle $(j, k, l)$ (\textbf{Fig.}\ref{ReichenbachConflict2}). Analogous to the previous one, in this example the arc $j \rightarrow k$ is spurious and the arc $k \rightarrow l$ should not be eliminated based on the vertices of the graph. That is, the graph underlying \textbf{Fig.}\ref{ReichenbachConflict2}(a) is \textbf{Fig.}\ref{ReichenbachConflict2}(b). Furthermore, $i$ is the common cause to $j$ and $k$ that 
causes the appearance of the arc $j \rightarrow k$. Again, if the triangle $(j, k, l)$ is visited first, the arc $k \rightarrow l$ can be erroneously eliminated for the same reasons, and then the arc $j \rightarrow k$ due to its spuriousness, obtaining the incorrect graph of \textbf{Fig.}\ref{ReichenbachConflict2}(c). 

\begin{figure}[ht]
	\centering
	\includegraphics[height=7cm]{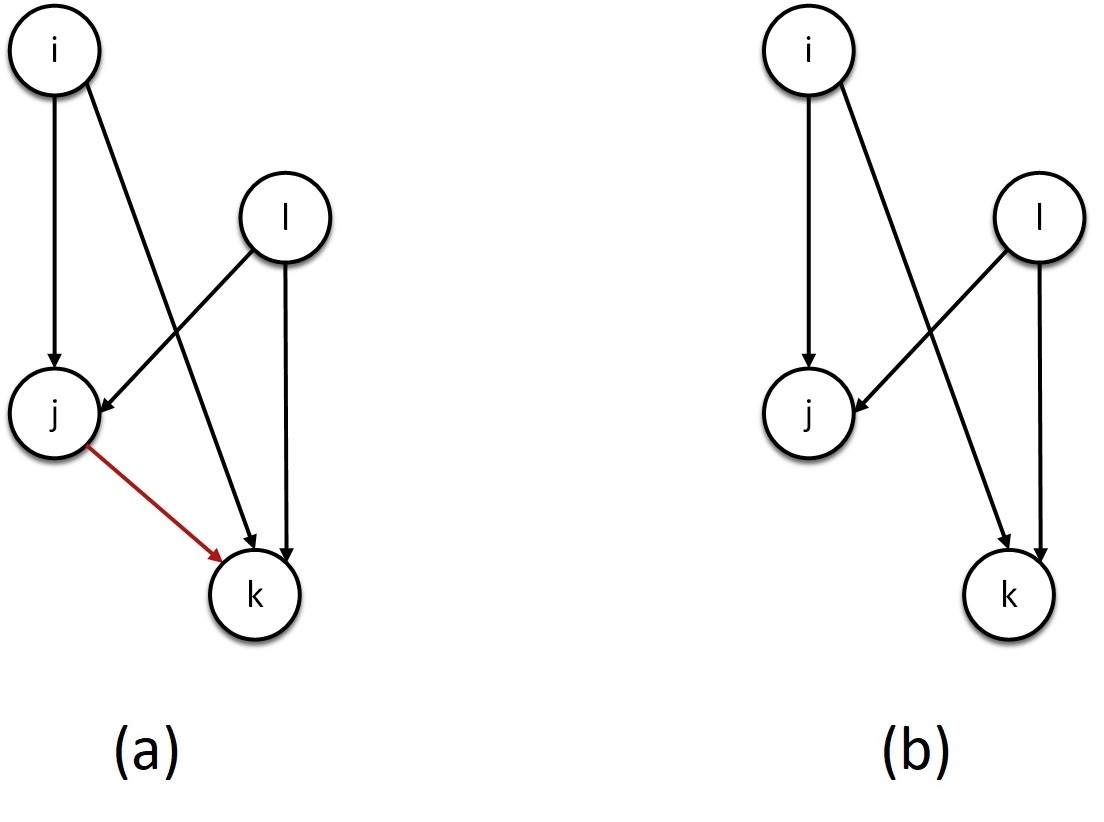}
	\caption{Third case of adjacent triangles, without conflicts for the elimination of spurious arcs}
	\label{ReichenbachNonConflict}
\end{figure}

\newpage 
In the third and last case, the arc $j \rightarrow k$ is the second leg of the triangles $(i, j, k)$ and $(l, j, k)$ (\textbf{Fig.}\ref{ReichenbachNonConflict}). By construction, the arc $j \rightarrow k$ is spurious in this example: the graph underlying \textbf{Fig.}\ref{ReichenbachNonConflict}(a) is \textbf{Fig.}\ref{ReichenbachNonConflict}(b), and i is the common cause to $j$ and $k$ that causes its appearance. In this case no conflicts are generated, since the only arc to eliminate in both triangles is the
spurious arc itself.

\newpage
The elimination of arcs that should not be removed based on the vertices of the
graph can be reduced by following an order criterion for the elimination based on the
average frequency of the ends of the arcs, from lowest to highest. In cases of equal
average frequency, there is no order criterion based on arc properties. In this scheme,
a heuristic but reasonable criterion is used: eliminating the arcs from highest to lowest
correlation. It is presumed that the external variables produce spurious relationships
of lower correlation than the internal variables, given that the latter have been
previously selected as the most relevant.

\section{Non-transitive paths}
\label{NonTransitivity}
Transitivity is a property of causal relations. In the deterministic case, it holds that
if \(i \rightarrow j\) and \(j \rightarrow k\), then \(i \rightarrow k\). Note that if we always have 
\(i \implies j\) and \(j \implies k\), then \(i \implies k\). In terms of \(H\), if \(H_{ij} = 1\) and
\(H_{jk} = 1\), then \(H_{ik} = 1\):

\begin{equation}
	\begin{aligned}
		&H_{ij} = 1 \implies \pi_{ij} = \pi_i\\
		&\pi_{ij} = \pi_i \implies \pi_{i \neg j} = 0\\
		&\pi_{i \neg j} = 0 \implies \pi_{i \neg j k} = 0\\
		&H_{jk} = 1 \implies \pi_{jk} = \pi_j\\
		&\pi_{jk} = \pi_j \implies \pi_{ijk} = \pi_{ij}\\
		&\pi_{ik} = \pi_{ijk} + \pi_{i \neg j k}\\
		&\pi_{ik} = \pi_{ij} + 0\\
		&\pi_{ik} = \pi_{ij}\\
		&\pi_{ik} = \pi_i\\
		&\pi_{ik} = \pi_i \implies H_{ik} = 1
	\end{aligned}
\end{equation}

But in the stochastic case these conditions are not met exactly, and transitivity may be lost. To
illustrate, see the following example that represents a possible distribution of three variables $i$, $j$, and $k$ in a set of 22 individuals:

\begin{table}[H]
	
	\centering	
	\resizebox{450pt}{50pt}{%
		
	\begin{tabular}{cccccccccccccccccc}
	\toprule
	& 1 & 2 & 3 & 4 & 5 & 6 & 7 & 8 & 9 & 10 & 11 & 12 & 13 & 14 & 15 & ... & 22 \\
	\midrule
	\midrule
	\(i\) & \textbf{1} & \textbf{1} & \textbf{1} & \textbf{1} & \textbf{1} & \textbf{1} & 0 & 0 & 0 & 0 & 0 & 0 & 0 & 0 & 0 & ... & 0 \\
	\(j\) & 0 & 0 & \textbf{1} & \textbf{1} & \textbf{1} & \textbf{1} & \textbf{1} & \textbf{1} & \textbf{1} & 0 & 0 & 0 & 0 & 0 & 0 & ... & 0 \\
	\(k\) & 0 & 0 & 0 & 0 & \textbf{1} & \textbf{1} & \textbf{1} & \textbf{1} & \textbf{1} & \textbf{1} & \textbf{1} & \textbf{1} & 0 & 0 & 0 & ... & 0 \\
	\bottomrule
	\end{tabular}
	}
	\caption{Possible distribution of three variables $i$, $j$, and $k$ in 22 individuals}
\end{table}

In principle, none of the implications \(i \implies j\) or \(j \implies k\) hold strictly, but the number of errors present in each one is small.

For \(H_0 = 0.5\) we have:

\begin{equation}
	\begin{aligned}
		&H_{ij} = \frac{\pi_{ij} - \pi_i\pi_j}{\pi_i(1 - \pi_j)} = \frac{\frac{4}{22} - \frac{6}{22}\frac{7}{22}}{\frac{6}{22}(1 - \frac{7}{22})} \approx 0.51 > H_0\\
		&H_{jk} = \frac{\pi_{jk} - \pi_j\pi_k}{\pi_j(1 - \pi_k)} = \frac{\frac{5}{22} - \frac{7}{22}\frac{8}{22}}{\frac{7}{22}(1 - \frac{8}{22})} \approx 0.55 > H_0\\
	\end{aligned}
\end{equation}

So, as $\pi_i < \pi_j < \pi_k$, \(i\) is a sufficient cause of \(j\), y \(j\) is a sufficient cause of \(k\) (assuming that there is no spuriousness). By transitivity, it must also be true that \(i\) is a sufficient cause of \(k\). Nevertheless:

\begin{equation}
	\begin{aligned}
&H_{ik} = \frac{\pi_{ik} - \pi_i\pi_k}{\pi_i(1 - \pi_k)} = \frac{\frac{2}{22} - \frac{6}{22}\frac{8}{22}}{\frac{6}{22}(1 - \frac{8}{22})} \approx -0.047 < H_0\\
	\end{aligned}
\end{equation}

That is, \(i\) is not a sufficient cause of \(k\). This does not constitute a violation 
of causal transitivity in the underlying deterministic case, since the following situation could occur:

\begin{figure}[ht]
	\centering
	\includegraphics[height=7cm]{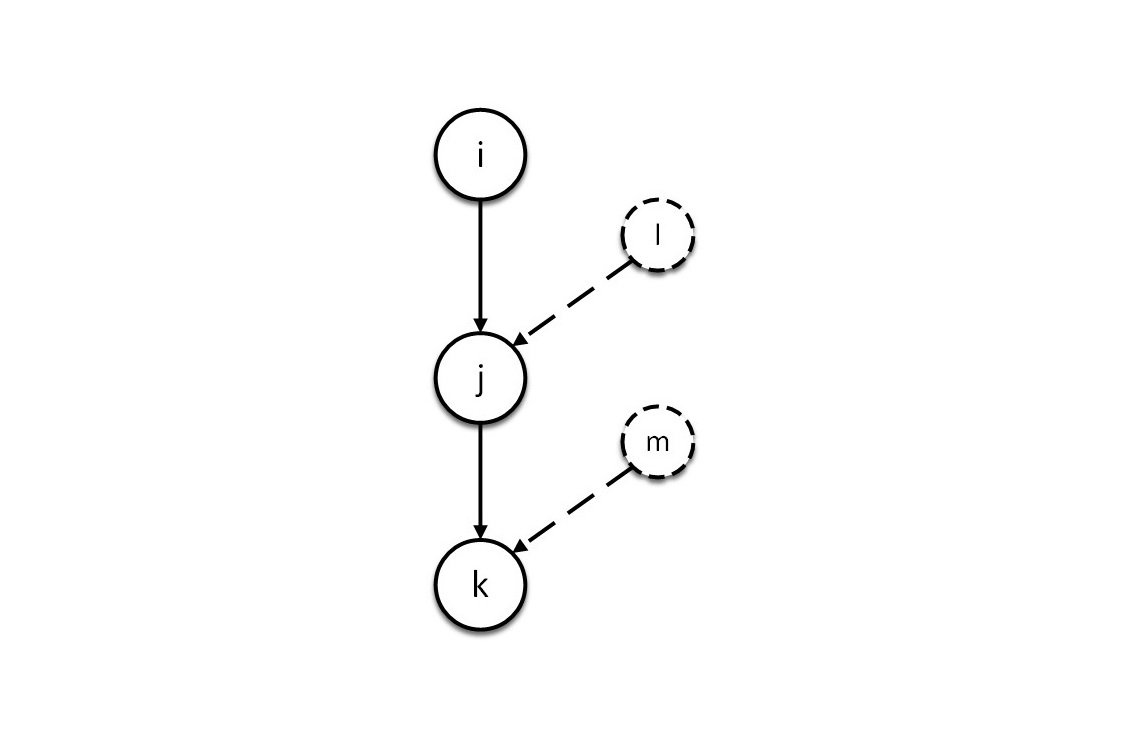}
	\caption{Example of possible latent contributory causes}
\end{figure}

In this example, $l$ contributes (together with $i$) in the production of effect $j$, and $m$ contributes
(together with $j$) in the production of effect $k$. Then in the causal relationship \(i \rightarrow k\) 
Two external causes contribute, losing causal sufficiency. It can be said that in reality the arcs \(i \rightarrow j\) and \(j\rightarrow k\) represent quasi-sufficient causal relations that emerge, for example, when contributory causes $l$ and $m$ have little variability in the data.

En este ejemplo, \(l\) contribuye (junto a \(i\)) en la producción del
efecto \(j\), y \(m\) contribuye (junto a \(j\)) en la producción del
efecto \(k\). Luego en la relación causal \(i \rightarrow k\)
contribuyen dos causas externas, perdiéndose la suficiencia causal.
Puede decirse que en realidad los arcos \(i \rightarrow j\) y
\(j\rightarrow k\) representan relaciones causales cuasi-suficientes que
emergen, por ejemplo, cuando las causas contributivas \(l\) y \(m\)
tienen poca variabilidad en los datos. In any case, one should not dispense 
with the non-transitive path $i \to j \rightarrow k$, since it provides 
useful information about the underlying causal relationships.

\section{Redundant arcs}
\label{Mokken}

See the following matrix, corresponding to a possible distribution of three variables $i$, $j$, and
$k$, in a set of 22 individuals:

\begin{table}[H]
	
	\centering	
	\resizebox{450pt}{50pt}{%
		
		\begin{tabular}{cccccccccccccccccc}
			\toprule
			& 1 & 2 & 3 & 4 & 5 & 6 & 7 & 8 & 9 & 10 & 11 & 12 & 13 & 14 & 15 & ...& 22 \\
			\midrule
			\midrule
			\(i\) & \textbf{1} & \textbf{1} & \textbf{1} & \textbf{1} & \textbf{1} &
			\textbf{1} & 0 & 0 & 0 & 0 & 0 & 0 & 0 & 0 & 0 & ... & 0 \\
			\(j\) & \textbf{1} & \textbf{1} & \textbf{1} & \textbf{1} & \textbf{1} &
			\textbf{1} & \textbf{1} & 0 & 0 & 0 & 0 & 0 & 0 & 0 & 0 & ... & 0 \\
			\(k\) & \textbf{1} & \textbf{1} & \textbf{1} & \textbf{1} & \textbf{1} &
			\textbf{1} & \textbf{1} & \textbf{1} & 0 & 0 & 0 & 0 & 0 & 0 & 0 & ... &
			0 \\
	\bottomrule
	\end{tabular}
	}
	\caption{Possible distribution of three variables $i$, $j$, and $k$, in 22 individuals}
\end{table}

As can be corroborated, the material implications \(i \implies j\) and \(j \implies k\) are strictly fulfilled.
Then \(H_{ij} = 1\), \(H_{jk} = 1\) and therefore \(H_{ik} = 1\). Since \(\pi_i < \pi_j < \pi_k\), 
the arcs \(i \rightarrow j\), \(j \rightarrow k\) and \(i \rightarrow k\) are considered to represent sufficient causal relations. However, in this triangle \((i, j, k)\) the hypotenuse does not provide new information regarding the legs: 
the causal relationship it represents is completely describable by the latter. Therefore, it is a
redundant arc. This is because \(i \rightarrow k\) actually represents an indirect causal relationship
from \(i\) to \(k\) through \(j\), a product of transitivity.

This particular example illustrates a deterministic case, and in these cases transitivity
always holds, as seen above. Therefore, to identify  \(i \rightarrow k\) as a transitive arc, it is enough
that \(i \rightarrow j\) and \(j \rightarrow k\) exist. But in the stochastic case this generally does not hold. To
illustrate, see the following example:

\begin{table}[H]
	
	\centering	
	\resizebox{450pt}{50pt}{%
		
		\begin{tabular}{cccccccccccccccccc}
		\toprule
		& 1 & 2 & 3 & 4 & 5 & 6 & 7 & 8 & 9 & 10 & 11 & 12 & 13 & 14 & 15 & ...
		& 22 \\
		\midrule
		\midrule
		\(i\) & \textbf{1} & \textbf{1} & \textbf{1} & \textbf{1} & \textbf{1} &
		\textbf{1} & 0 & 0 & 0 & 0 & 0 & 0 & 0 & 0 & 0 & ... & 0 \\
		\(j\) & \textbf{1} & \textbf{1} & \textbf{1} & \textbf{1} & \textbf{1} &
		0 & \textbf{1} & \textbf{1} & 0 & 0 & 0 & 0 & 0 & 0 & 0 & ... & 0 \\
		\(k\) & \textbf{1} & \textbf{1} & \textbf{1} & \textbf{1} & \textbf{1} &
		\textbf{1} & 0 & 0 & \textbf{1} & \textbf{1} & 0 & 0 & 0 & 0 & 0 & ... &
		0 \\
	\bottomrule
	\end{tabular}
	}
	\caption{Possible distribution of three variables $i$, $j$, and $k$ in 22 individuals}
	\label{NoSimplification}	
\end{table}

\newpage
For \(H_0 = 0.5\) we have:

\begin{equation}
	\begin{aligned}
		&H_{ij} = \frac{\pi_{ij} - \pi_i\pi_j}{\pi_i(1 - \pi_j)} = \frac{\frac{5}{22} - \frac{6}{22}\frac{7}{22}}{\frac{6}{22}(1 - \frac{7}{22})} \approx 0.75 > H_0\\
		&H_{jk} = \frac{\pi_{jk} - \pi_j\pi_k}{\pi_j(1 - \pi_k)} = \frac{\frac{5}{22} - \frac{7}{22}\frac{8}{22}}{\frac{7}{22}(1 - \frac{8}{22})} \approx 0.55 > H_0\\
		&H_{ik} = \frac{\pi_{ik} - \pi_i\pi_k}{\pi_i(1 - \pi_k)} = \frac{\frac{6}{22} - \frac{6}{22}\frac{8}{22}}{\frac{6}{22}(1 - \frac{8}{22})} = 1 > H_0
	\end{aligned}
\end{equation}

Therefore, as $\pi_i < \pi_j < \pi_k$, it is considered that the arcs \(i \rightarrow j\), \(j \rightarrow k\) and \(i \rightarrow k\) represent sufficient causal relations. In the causal sufficiency relation \(i \rightarrow k\) there are no errors. On the other hand, in the causal chain \(i \rightarrow j \rightarrow k\) there are a total of $3$ errors, corresponding to columns $6$-$8$ of the matrix. Given this situation, in the triangle \((i, j, k)\) it is not possible to explain, using the legs , the causal relationship represented by the hypotenuse. Therefore, either \(i \rightarrow k\) is a direct causal relationship in itself, or it is the union of a direct and an indirect relationship, or it represents another indirect relationship in which $j$ does not intervene.

A criterion is necessary, then, to identify transitivity relations of which not, with the aim of eliminating all the redundancies of $G$.

Starting from the first example (deterministic case of causality), it is observed that the transitive
relation \(i \rightarrow k\) meets the following conditions:

\begin{equation}
		\pi_{ij} = \pi_{ik} \leq \pi_{jk}\\
\end{equation}
\begin{equation}
		\pi_{ \neg i \neg j} \geq \pi_{\neg i \neg k} = \pi_{\neg j \neg k}
\end{equation}

On the other hand, in case of statistical independence between \(i\), \(j\) and $k$ two by two (case of nonexistence of causal relationships), the conditions are met:

\begin{equation}
		\pi_i\pi_j < \pi_i\pi_k < \pi_j\pi_k \implies \pi_{ij} < \pi_{ik} < \pi_{jk}
\end{equation}
\begin{equation}
		\pi_{\neg i}\pi_{\neg j} > \pi_{\neg i}\pi_{\neg k} > \pi_{\neg j}\pi_{\neg k} \implies \pi_{\neg i \neg j} > \pi_{\neg i \neg k} > \pi_{\neg j \neg k}
\end{equation}

In the intermediate case, transitive relations are expected to fulfill a mixed condition of both. That is: 

\begin{equation}
	\pi_{ij} \leq \pi_{ik} \leq \pi_{jk}
	\label{Mokken1}
\end{equation}
\begin{equation}
	\pi_{\neg i \neg j} \geq \pi_{\neg i \neg k} \geq \pi_{\neg j \neg k}
	\label{Mokken2}
\end{equation}

These inequalities were rigorously deduced by Mokken within the framework of a latent
variable theory.\brackcite{Mokken} Even so, it was preferred to avoid the formalization of Mokken for two
fundamental reasons: \textbf{1)} Mokken does not address the issue of causality (and therefore, neither
that of indirect causality) and it is only a reinterpretation of its result that we allows it to be used
in our context, and \textbf{2)} Mokken's conceptual apparatus requires a series of assumptions that,
although compatible with the case of interest, require a much more detailed treatment, which
is beyond the scope of this thesis. We refer the reader to Mokken's book, under the assumption
that \textit{prima facie} the Loevinger coefficient is capable of measuring the validity of causal inference.

Therefore, for every triangle \((i, j, k)\), the hypotenuse is considered to represent an indirect
causal relationship described by the legs if the proposed conditions are met. In this case, the
hypotenuse \(i \rightarrow k\) of $G$ is eliminated.

Note that in the example related to table \ref{NoSimplification}  condition \ref{Mokken1} is not met, since $\pi_{ik} > \pi_{jk}$. Therefore, it is correctly identified that the arc $i \rightarrow k$ is not transitive.

Finally, if any of the legs of \((i, j, k)\) is an undirected edge, it is directed in the sense of the
identified transitive causal relationship. For example, by identifying an arc \(i \rightarrow k\) as the indirect
causal relation \(i \rightarrow j \rightarrow k\), the edge between $i$ and $j$ is directed as if undirected. Analogously for
the edge between $j$ and $k$.

\begin{figure}[ht]
	\centering
	\includegraphics[height=9cm]{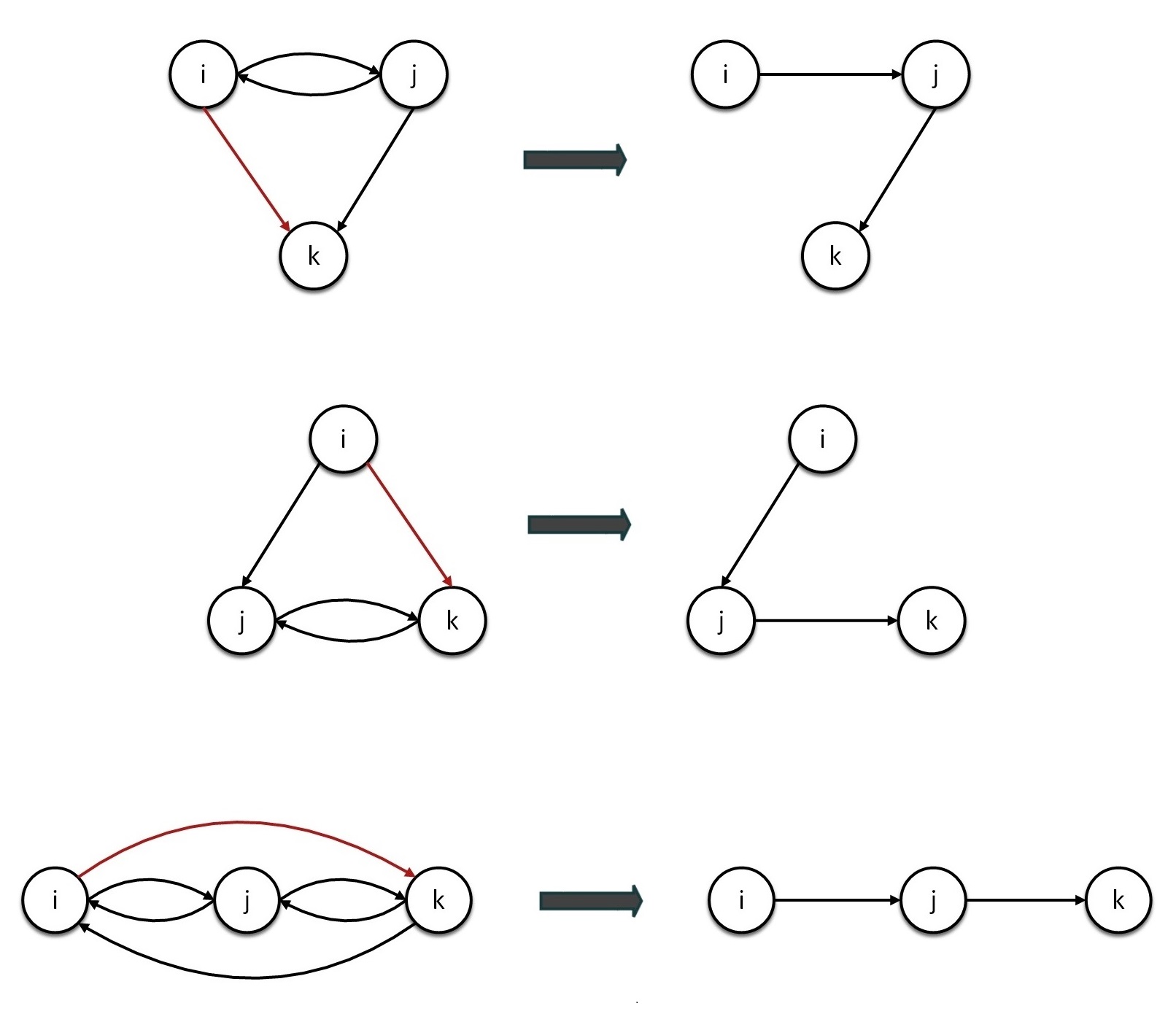}
\caption{Examples of undirected edge orientation in the presence of transitivity}
\end{figure}

\subsection{Order of elimination of redundant arcs}
\label{TransitiveOrder}
Removing arcs by this method must follow a specific order, to avoid conflicts. Three
cases may arise: when the transitive hypotenuse of the triangle considered is the transitive
hypotenuse of another triangle, and when it is the first or second leg of another transitive
triangle. In the first case, the hypotenuse could be eliminated by either of the two triangles
involved. However, in the second and third cases, eliminating the hypotenuse destroys
another transitive triangle. The cases in which the transitive hypotenuse is the first or second
leg of another transitive triangle are, essentially, identical, so only one of them will be
addressed below.

\begin{figure}[ht]
	\centering
	\includegraphics[height=6.5cm]{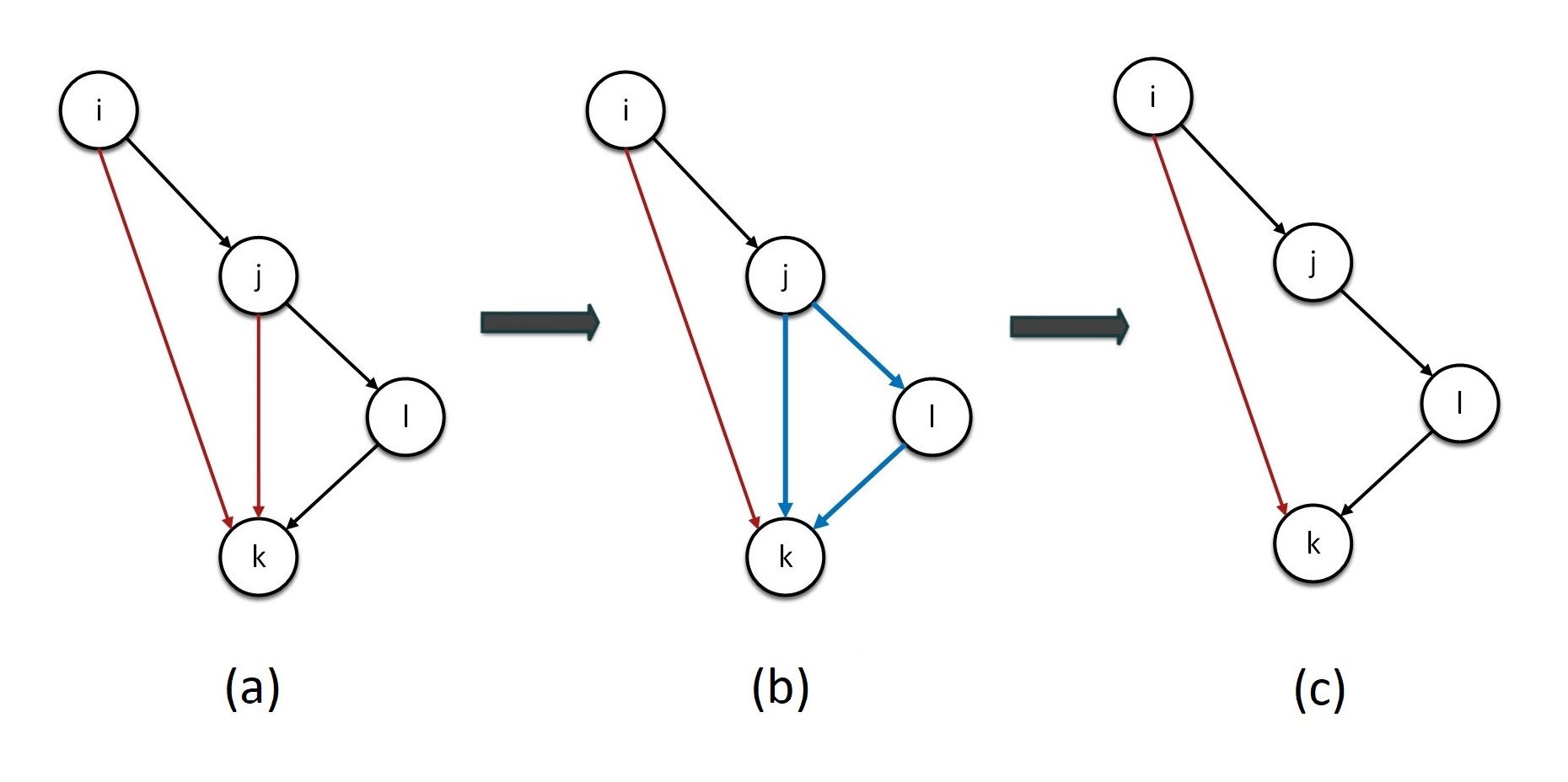}
	\caption{Incorrect removal order for transitive arcs}
	\label{MokkenConflict}
\end{figure}

Take as an example the case in which the transitive hypotenuse of a triangle is the
second leg of another transitive triangle (\textbf{Fig.}\ref{MokkenConflict}). In this case, 
the hypotenuse of the triangle \((j, l, k)\) is the second leg of the triangle \((i, j, k)\). 
By construction, the arcs \(i \rightarrow k\) and \(j \rightarrow k\)
represent the indirect causal relations  \(i \rightarrow j \rightarrow k\) and \(j \rightarrow l \rightarrow k\) respectively. By choosing the second triangle first (\textbf{Fig.}\ref{MokkenConflict}(b)) and eliminating \(j \rightarrow k\), the first triangle disappears. Therefore, it is impossible to eliminate the arc \(i \rightarrow k\) with the proposed method.

\begin{figure}[ht]
	\centering
	\includegraphics[height=6.5cm]{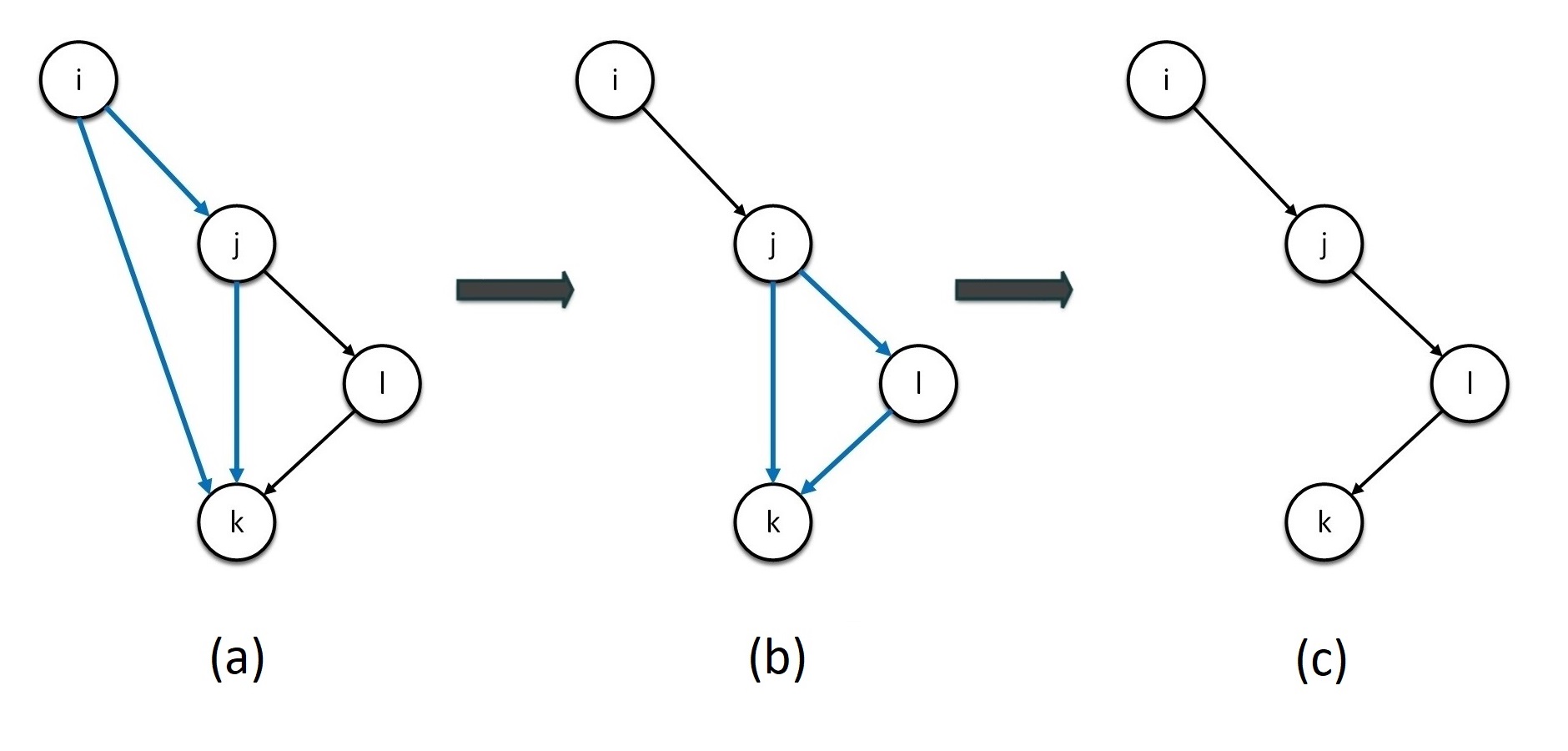}
	\caption{Correct removal order for transitive arcs}
	\label{MokkenNonConflict}
\end{figure}

On the other hand, if the first triangle is chosen first (\textbf{Fig.}\ref{MokkenNonConflict}(a)) and the arc \(i \rightarrow k\) is eliminated , the second triangle can later be visited (\textbf{Fig.}\ref{MokkenNonConflict}(b)) and \(j \rightarrow k\) eliminated without problems, obtaining the correct underlying graph (\textbf{Fig.}\ref{MokkenNonConflict}(c)).

\newpage

Note that in general, when the transitive hypotenuse is the first or second leg of another
transitive triangle, it is true that said hypotenuse has less than or equal length than that of the other triangle. Since the hypotenuse of the other triangle must be removed first, if they are removed in order from longest to shortest by length, no conflicts as long as the lengths are different. That is, an arc \(i \rightarrow j\) will be examined before another \(l \rightarrow k\) if \(|\pi_i - \pi_j| > |\pi_l - \pi_k|\).

However, under this criterion, the order of elimination is arbitrary in cases of equal length. Only when two triangles are adjacent, and their hypotenuses have equal length, an order must be established for the discarding of redundant arcs. In these conditions, three cases are distinguished, according to the role played the hypotenuse of the triangle considered in the adjacent triangle (first leg, second leg, and hypotenuse). In the case of the hypotenuse, the order of elimination is arbitrary. Eliminating the hypotenuse of a transitive triangle, with the help of another transitive triangle that also has it as hypotenuse means, in any way, eliminating it. In the case where the arc in question is the second leg of the adjacent triangle, the order of elimination can be found in terms of Mokken's inequalities, that is, by ordering the arcs by frequency of coincidences from smallest to largest. However, this order fails in the case where the arc is the first leg of the adjacent triangle where the order of elimination would be dictated by the frequency of coincidences, but in reverse, from highest to lowest.

In the algorithm under discussion, one of these orders is taken by frequencies of coincidences, knowing that there may be redundant arcs that will not be eliminated. In any case, these cases must be rare as they must satisfy three conditions: \textbf{1)} that for a transitive triangle there exists at least one other adjacent triangle, \textbf{1)} transitive triangle, \textbf{2)} transitive triangle transitive, \textbf{2)} that the hypotenuses of the adjacent triangles be of equal length, and \textbf{3)} that the hypotenuses of the adjacent triangles be of equal length. equal length, and \textbf{3)} that one of the triangles takes the redundant arc of the other as its first leg. of the other as the first leg.

A simple algorithm for solving this problem would be to postpone the elimination of arcs of equal length until we move to a different length in the order of arcs. to a different length in the order of arcs. Unfortunately, this leads to a large computational computational cost since there is no way to distinguish a priori whether the arc of equal arc of equal length will be hypotenuse, first or second leg of a transitive triangle. transitive triangle.

We then resort to the same heuristic criterion used as in the heading \ref{Reichenbach_order}: order the arcs of equal length from greatest to least correlation. correlation. The idea is to give priority to the hypotenuses with the highest correlation between their extremes, since they are the ones that presumably represent direct and non-redundant causal relationships.

\section{Remaining undirected edges}
\label{Robustness}

The methods used to eliminate spurious and redundant arcs can partially solve the
problem of undirected edges in $G$. In specific cases, they allow them to be completely removed
or their real direction determined. However, they are not sufficient solutions: the first requires
the spuriousness of the edge to eliminate it, while the second requires the presence of
transitivity in the triangle under analysis to eliminate the edge or orient it according to the case.
Consequently, other approaches are necessary to orient or eliminate possible remaining undirected edges.

In an initial approach, for each undirected edge \(i \leftrightarrow j\) it is possible to discern cause from
effect using the asymmetric properties of causality. For example, from a property of $i$ inherent
to causes and improper to effects, one could conclude that the causal relation underlying \(i \leftrightarrow j\)
is \(i \rightarrow j\), and eliminate the opposite arc \(j \rightarrow i\). Among the possible causality asymmetries to
analyze for this purpose, one of the best candidates is robustness.

Hausman defines robustness as an asymmetric property of causality, which expresses the
invariance of the cause-effect relationship with respect to the frequency of the cause or how it
arises, but not with respect to the frequency of the effect or how it arises.\brackcite{Hausman}
Modifying the frequency of a cause should not break the link between it and the effect.
However, due to the necessary condition \(\pi_i < \pi_j\) , increasing the frequency of a cause $i$ 
is not always feasible (an increase in the frequency of $i$ could violate the inequality). On the other
hand, decreasing the frequency of a cause $i$ does not cause conflicts in terms of robustness:
the necessary condition remains true, and the number of errors in \(i \rightarrow j\) can only be maintained
or decreased.

This criterion can be used to orient undirected edges. For a given arc \(i \leftrightarrow j\), the robustness
of the relations \(i \rightarrow j\) and \(j \rightarrow i\) is checked. If the first is robust (a decrease in the frequency of $i$ keeps the cause-effect relationship from $i$ to $j$ intact), while the second does not, then it is assumed that \(i \rightarrow j\) is the causal relationship underlying \(i \leftrightarrow j\), and the arc \(j \rightarrow i\) is eliminated. Analogously otherwise. 

At this point, the question is how to correctly alter the frequency of a variable. Decreasing the
frequency of a variable $i$ does not only alter the value \(\pi_i\), it also varies the individuals in which the
variable $i$ takes a value of $1$. This redistribution of values of $i$ cannot be arbitrary, as it could change
the nature of the variable, invalidating the analysis.

In a stochastic case, the following solution can be taken to obtain a valid redistribution of a variable $i$:

\begin{itemize}
\item Carry out a new survey of variable \(i\) on the same population of individuals.
\item Calculate the intersection between the results of the original survey and this new survey.
\item Let \(ii\)  be this intersection, take \(ii\) as the new variable \(i\).
\end{itemize}

This solution is justified in that $ii$ takes a value of $1$ only in individuals in which $i$ is $1$ for both
surveys, and therefore collects the most consistent information about $i$. On the other hand, \(p_{ii} \leq p_i\), so the desired decrease in frequency is expected to be achieved.

With the new distribution $ii$, we calculate \(H_{iij} = \frac{\pi_{iij} - \pi_{ii}\pi_j}{\pi_{ii}(1 - \pi_j)}\) 
to check if it remains above the validity threshold $H_0$ (note that the expression $H_{iij}$ is obtained from $H_{ij}$ 
by replacing all the values referring to $i$ with those of $ii$). If $H_{iij} > H_0$ is true, then the relationship \(i \rightarrow j\) is robust. If by proceeding analogously with $j$, it is confirmed that \(j \rightarrow i\) is not robust \(H_{jji} < H_0\), then it is assumed that \(i \rightarrow j\) is the causal relationship underlying \(i \leftrightarrow j\), and the arc \(j \rightarrow i\) is eliminated. In cases where
both relationships are robust, the edge can be oriented based on the values \(H_{iij}\) and \(H_{jji}\). If \(H_{iij} > H_{jji}\) is true , it makes more sense to infer causality from $i$ to $j$ than vice versa, given its greater
validity. Analogously for the case \(H_{iij} < H_{jji}\). The cases in which \(H_{iij} = H_{jji}\) hold should be
considerably few or non-existent.

Even if the problem seems solved, conducting a new survey for a variable $i$ is usually
impractical or impossible. An alternative then is approximating $ii$, which in practice is reduced to
approximating the values\(\pi_{ii}\) and \(\pi_{iij}\) that are used to calculate \(H_{iij}\).

Let \(i-1\) be the parent of \(i\) with the highest frequency of matches with \(i\), \(i-1\) can be taken as
the approximation of a new survey of $i$. However, considering the values \(\pi_{i,i-1}\) and \(\pi_{ii}\) approximate
may not be rigorous enough. 

However, it can be expected that the similarities between both terms are maintained even in cases of
independence, due to the way in which \(i-1\) was chosen, ie, \(\frac{\pi_{ii}}{\pi_{i, i-1}} \approx \frac{\pi_{i}^2}{\pi_i\pi_{i-1}}\). Solving, we obtain \(\pi_{ii} \approx \frac{\pi_i\pi_{i, i-1}}{\pi_{i-1}}\) as an approximate expression for \(\pi_{ii}\). On the other hand, let \(i+1\) be the child of $i$ with the highest frequency of
coincidences with $i$, the expression \(\pi_{ii} \approx \frac{\pi_i\pi_{i, i+1}}{\pi_{i+1}}\) is obtained in an analogous way as an approximation of \(\pi_{ii}\). These two expressions were proposed by Mokken within the framework of his theory\brackcite{Mokken}, also as an approximation of the coincidences between surveys of the same variable, with similar definitions of \(i-1\) and \(i+1\).

Using reasoning similar to the previous one, it is expected that the similarities between \(\pi_{ij}\) and \(\pi_{iij}\) will
remain proportional in case of independence between $i$ and $j$, this is, \(\frac{\pi_{ij}}{\pi_{iij}} \approx \frac{\pi_i\pi_j}{\pi_{ii}\pi_j}\). Solving, we obtain \(\pi_{iij} \approx \frac{\pi_{ii}\pi_{ij}}{\pi_i}\) as an approximate expression for \(\pi_{iij}\).

By approximating the values \(\pi_{ii}\), \(\pi_{iij}\), \(\pi_{jj}\) and \(\pi_{jji}\) in this way, it is possible to calculate \(H_{iij}\) and \(H_{jji}\) to carry out the rest of the proposed analyses. In practice, for each undirected edge \(i \leftrightarrow j\) the values are approximated using \(i-1\) and \(j-1\) first and, if this is not sufficient \(H_{iij} = H_{jji}\)), using \(i+1\) and \(j+1\) later.

In the small cases in which \(H_{iij} = H_{jji}\) is met, one last criterion is applied: calculate and compare the
values \(H_{iijj}\) and \(H_{jjii}\). \(H_{iijj}\) is a measure of the validity of the inference \(i \rightarrow j\) considering only the most consistent information of $i$ and $j$.
Therefore, if \(H_{iijj} > H_{jjii}\) then it makes more sense to infer causality from $i$ to $j$ than vice versa, given its
greater validity. Analogously for H\(H_{iijj} < H_{jjii}\). On the other hand, if \(H_{iijj} \leq H_0\) and \(H_{jjii} \leq H_0\) are true, it is considered that between $i$ and $j$ there is actually no causal sufficiency relationship: since causal sufficiency is lost in both directions when considering only the most consistent data for each variable, then it can be
concluded that such sufficiency is spurious, a product of noise in the data.

Since the values \(\pi_{ii}\) and \(\pi_{jj}\), to calculate
\(H_{iijj} = \frac{\pi_{iijj} - \pi_{ii}\pi_{jj}}{\pi_{ii}(1 - \pi_{jj)}}\)
it's just necessary to calculate \(\pi_{iijj}\). Applying the same idea as with \(\pi_{iij}\), it's assumed
\(\frac{\pi_{ij}}{\pi_{iijj}} \approx \frac{\pi_i\pi_j}{\pi_{ii}\pi_{jj}}\),
and \(\pi_{iijj} \approx \frac{\pi_{ii}\pi_{jj}\pi_{ij}}{\pi_i\pi_j}\) is obtained as  an approximate expression for \(\pi_{iijj}\).

\chapter{Implementation details and experiments}\label{chapter:implementation}

The methodology for causal discovery proposed in the previous chapter is original,
and is presented for the first time in this document. For its implementation, a code for
scientific use called \textbf{CChains} (for causal chains, in English) is developed in the C++
programming language. The latter is chosen because it is a high-speed compiled
language, ideal for the intense calculation required by the algorithm.

The program consists of several phases, which correspond to each of the steps of
the methodology. First, the causal graph is constructed {[}phase 1{]}, using the metric \(H_{ij}\)
for each pair of vertices and the threshold \(H_0\). Second, we proceed to simplify the
causal graph {[}phase 2{]}, by eliminating arcs. This phase is further subdivided into three
stages dedicated to solving the problems of spurious arcs (due to a common cause)
{[}stage 1{]}, redundant (due to transitivity) {[}stage 2{]} and remaining undirected edges
{[}stage 3{]}, in that order. In each stage, all and only the arcs of the simplified graph in the
previous stage or phase are consulted. All phases and stages are independent, in the
sense that they consist of their own inputs and outputs, and are associated with different
modules.

In the following, the input, output and operating specifications of each phase and stage are discussed, 
after an explanation of the basic general configuration.

\section{General configuration}

The program code base is available in the public GitHub repository \texttt{https://github} \texttt{.com/jean-pierre-gm/CChains}. It can be downloaded through the console by using the command \texttt{git clone https://github.com/jean-pierre-gm/CChains} and, if you have Cmake installed, compiled using the command \texttt{cmake} from the downloaded directory, to obtain the corresponding executable. Since the program is intended for scientific use, it does not have a graphical user interface. However, the configuration of the program is extremely simple, as will be described below. 

The configuration is done by means of a plain text file. The address of this file is the only argument that the program receives, e.g:

\begin{lstlisting}
	C:\Users\Jean\Desktop> CChains.exe parameters.txt
\end{lstlisting}

This file specifies the address of the remaining input (data) and output files, among
other specifications. The configuration file has the next format, e.g,

\begin{lstlisting}
	run_mode=build
	
	metric=H
	threshold=0.5
	
	graph_builder_in=sample.txt
	graph_builder_out=graph_file.txt
	graph_builder_frequencies_out=frequencies.txt
	
	run_summary=out.txt
\end{lstlisting}

Each line of the configuration file must correspond to a single input variable, following the format texttt{$<$variable name$>$=$<$value$>$}. Any line that does not respect this format is taken as a comment and is therefore not processed as input. 

In particular, the input variables \texttt{run\_mode}, \texttt{metric}, \texttt{threshold} and \texttt{run} \newline \texttt{summary} are global.  

The variable \texttt{metric} refers to the measure associated to each pair of vertices to check whether or not an arc exists between them (in phase 1), and on the basis of which spurious arcs and undirected edges are simplified (in phase 2, stages 1 and 3, respectively). Although the CChains methodology starts from $H$, it may be of interest to analyze the behavior of some phase or stage of the algorithm (in particular, the construction of the graph) using other metrics. The metrics available are the Loevinger coefficient (\texttt{metric=H}), the Pearson correlation coefficient (\texttt{metric=r}) and the two measures of probabilistic causality explored under the heading \ref{ProbCausation} (\texttt{metric=pc1} and \texttt{metric=pc2}). This input variable is mandatory.

The variable \texttt{threshold} refers to the threshold that must be exceeded by the considered metric to determine the presence of an arc between two vertices (in phase 1), as well as to simplify spurious arcs and undirected edges (in phase 2, stages 1 and 3, respectively). It corresponds to the parameter \(H_0\) in the case \texttt{metric=H}. This input variable is mandatory.

La variable  \texttt{run\_mode} especifica qué fases o etapas del programa se desean
ejecutar. En particular, la instrucción

\begin{itemize}
	\item
	\texttt{run\_mode=build} is used to build the causal graph (phase 1),
	\item
	\texttt{run\_mode=Reichenbach} to eliminate spurious arcs (phase 2, stage 1),
	\item
	\texttt{run\_mode=Mokken} to remove redundant arcs (phase 2, stage 2),
	\item
	\texttt{run\_mode=Robustness} to remove remaining undirected edges (phase 2, stage 3).
\end{itemize}

The variables corresponding to stages 1 and 2 of phase 2 are named \textbf{Reichenbach} and \textbf{Mokken} because Reichenbach's principle of common cause (heading \ref{ProbCausation}), and the inequalities \ref{Mokken1} and \ref{Mokken2} proposed by Mokken, are the main inspiration for these phases of the proposed algorithm.   

The phases and stages can be chained with the operator \texttt{\&}, to be executed one after the previous one. executed one after the previous one, e.g, \texttt{run\_mode=build\&} \texttt{Reichenbach\_test\&Robustness\_test}. This input variable is mandatory.

The variable \texttt{run summary} determines the address of the file to which the run report for each phase is written. the execution report of each phase or stage is written to. The report contains summary information about the number of arcs before and after each stage or phase. of each stage or phase. This input variable is optional. In its absence, the program writes the output report to the console.

\section{Construction of the causal graph}

The program constructs a causal graph \(G=<V,E>\), from a matrix \(M\) of binary variables \(V\) by individuals \(I\), using the method proposed in the section \ref{Build}. The matrix data is received as a plain text file, the address of which is determined by the value of the input variable in the configuration file \texttt{graph\_builder\_in} (e.g., \texttt{graph\_builder\_in=sample.txt}). 
This file contains the matrix \(M\) in tabular form, where each row represents a variable from \(V\) and each column represents an individual from \(I\). The number of rows and columns matches the number of individuals in the matrix. The number of rows and columns coincides with \(|V|\) and \(|I|\), respectively. The supported values of each component are $0$ or $1$, and each column must be separated by \emph{normal} or tab (or \emph{tab}) spaces. Below is an example of a file with the correct formatting:

\begin{lstlisting}
	0	1	1	0	0	0	0	0
	0	1	1	1	1	1	1	1
	0	0	0	1	1	1	1	1
	0	1	0	1	1	1	1	1
	0	0	0	1	0	1	1	0
	1	1	1	1	1	1	1	0
\end{lstlisting}

So far, the only input data required for phase 1. As output, the program returns a file with the causal network, in the form of adjacency lists. adjacency lists. The variable \texttt{graph\_builder\_out} determines the output file of the network (e.g., \texttt{graph\_builder\_out=graph\_file.txt}) in plain text. The row \(i\) of this file corresponds to the adjacency list of the vertex \(i\in G\). In each column, the adjacencies of \(i\) are represented as pairs ($j$, $j$, $j$). are represented as pairs ($j$, $\pi_{ij}$), where the child index and the frequency of parent-child matches are separated by a space. In contrast, the columns are separated by a \emph{tab}. Finally, empty rows empty rows correspond to vertices without children. See the following example of output file for the network:

\begin{lstlisting}
	1 0.25		5 0.25	
	
	1 0.625		3 0.625	
	1 0.75	
	1 0.375		2 0.375		3 0.375		5 0.375	

\end{lstlisting}

These formats for graphs and data matrices will be used hereafter to describe the inputs and outputs of the algorithm.

Additionally, the input variable \texttt{graph\_builder\_frequencies\_out} determines the address of the output file for the frequency of the vertices of \(G\) (e.g., \texttt{graph\_builder} \texttt{\_frequencies\_out=frequencies.txt}). This entry is optional, and in its absence the frequencies of each vertex are not exported to file. When generated, this file contains one row and one column for each vertex. The order of rows is not in correspondence with the order of frequencies, but with the row index of the matrix \(M\) and of vertices in the graph \(G\).

\subsection{Input matrix compression}
\label{Compress}
The matrix may contain identical rows due to the finiteness of the sample of individuals and/or the origin of the binary matrix. In case identical rows are present, the associated variables are described by a single vertex. For this reason, from the initial set \(V\) of variables (rows of $M$), another final set \(V\) of vertices of $G$) must be constructed, such that to each variable of $V$ corresponds one and only one vertex of $V$, and to two variables of $V$ corresponds the same vertex of $V$, and to two variables of $V$ corresponds the same vertex of $V$, such that to each variable of $V$ corresponds the same vertex of $V$, such that to each variable of $V$ corresponds the same vertex of $V$, if and only if the corresponding rows of the matrix of the matrix $M$ are identical.

This process is called matrix compression and takes place in stage 1 of phase 1 (stage 1), and is specified in the input variable \texttt{run\_mode} by means of the value \texttt{sample\_compressor}. This C++ method (hereafter, \emph{method} in italics) receives the original matrix $M$, performs the compression process, and returns the resulting matrix \(M'\), where all rows are different from each other. The addresses of the input and output files of the matrix must be specified in the configuration file, by means of the variables \texttt{sample\_compressor\_in} and
\texttt{sample\_compressor\_out}.
	
The \emph{method} also returns another file, where each row corresponds to a vertex of \(V'\), and each column corresponds to the variables in \(V\) variables associated to it. The purpose of this file is to serve as a map between the variables and the vertices (or nodes) of the network. The address of this output file is specified by the input variable \texttt{sample\_compressor\_nodes\_out}.

The simplest run using matrix compression is set up as follows, e.g.:

\begin{lstlisting}
	run_mode=sample_compressor\&build
	
	metric=H
	threshold=0.5
	
	sample_compressor_in=sample.txt
	sample_compressor_out=compressed_sample.txt
	sample_compressor_nodes_out=nodes.txt
	
	graph_builder_in=sample.txt
	graph_builder_out=graph_file.txt
	graph_builder_frequencies_out=frequencies.txt
\end{lstlisting}

The matrix \(M'\) obtained from the compression process is the one to be used from now on in the rest of the phases of the algorithm, in particular used in the remaining phases of the algorithm, in particular in the construction of the network (phase 1). in the construction of the network (phase 1). It will always be possible to obtain, from a vertex of from a vertex of $V$, the associated variables of $V$ using the map between sets. using the map between sets.

\section{Causal graph simplification}

La fase 2 (simplificación del grafo) recibe un archivo de texto plano
con la descripción de un grafo \(G\) (listas de adyacencia, ver más
arriba), y un archivo con la matriz \(M\) asociada. Devuelve un archivo
de texto plano con el grafo \(G\) tras eliminar todos sus arcos
espurios, redundantes, y eliminar u orientar las aristas no dirigidas
remanentes, dependiendo de la etapa, y según los métodos propuestos en los
epígrafes \ref{Reichenbach}, \ref{Mokken}, y \ref{Robustness}, respectivamente.

Los archivos de entrada del grafo (a simplificar) y la matriz de muestra
deben especificarse como valores de la variables
\texttt{graph\_simplifier\_in} y \texttt{graph\_simplifier\_} \newline \texttt{sample\_in}, respectivamente. El archivo de salida del grafo (simplificado) de salida en \texttt{graph\_simplifier\_out}.

Un ejemplo de configuración puede ser el siguiente:

\begin{lstlisting}
	run_mode=Reichenbach&Mokken&Robustness
	
	metric=H
	threshold=0.5
	
	graph_simplifier_in=graph_file.txt
	graph_simplifier_sample_in=sample.txt
	graph_simplifier_out=simplified_graph_file.txt
	graph_simplifier_frequencies_in=frequencies.txt
\end{lstlisting}

The input variable \texttt{graph\_builder\_frequencies\_in} is optional, and specifies the input file for the vertex frequency listing. If not specified, the listing is recalculated from the input matrix.

If the chaining of more than one stage of phase 2 is specified in \texttt{run\_mode} (e.g., using \texttt{run\_mode=Reichenbach\&Mokken}), then the output network of one stage corresponds to the input network of the next stage. In that case, the output file corresponds to the sequence of stages. By default, the network resulting from any of the intermediate stages is not available at the end of the calculation. If necessary, the network resulting from a specific stage can be obtained by assigning the address of an output file to the variable named \textless{[}value of run\_mode corresponding to stage 2{]}\_out\textgreater{}(e.g., \texttt{Reichenbach\_out=reichenbach\_graph} \newline \texttt{\_file.txt} para la etapa 1 de la fase 2) for stage 1 of phase 2).

\section{Modules and dependencies}

The modules of the \emph{CChains} program are listed below, along with a brief description of its content:

\begin{itemize}
	\item
	\emph{\textbf{Definitions}}: type definitions and class declarations that are used in the rest of the
	modules. It has an \textbf{iomanager} submodule in which the \emph{methods} to manage the program's input 
	and output files are defined. 

	\item
	\emph{\textbf{Metrics}}: \emph{methods} for calculating correlation or causality measures (Loevinger's
	coefficient, Pearson's coefficient, and probabilistic causality measures), as well as
	their dependencies (\emph{methods} for calculating frequencies, frequencies of coincidences,
	among others)
	
	\item
	\emph{\textbf{Sample compressor}}: $M$ matrix compression \emph{method}.
	
	\item
	\emph{\textbf{Graph auxiliary methods}}: auxiliary \emph{methods} for managing graphs, eg, the \emph{method}
	to obtain the arcs of the graph, to obtain the undirected edges, to eliminate arcs, to
	transpose the graph, among others.
	
	\item
	\emph{\textbf{Graph builder}}: causal graph $G$ construction \emph{method}.
	
	\item
	\emph{\textbf{Graph simplifier}}: causal graph simplification modules.
	
	\begin{itemize}
		\item
		\emph{\textbf{Reichenbach test}}: \emph{method} for eliminating spurious arcs.
		
		\item
		\emph{\textbf{Mokken test}}: \emph{method} for eliminating redundant arcs.
		
		\item
		\emph{\textbf{Robustness test}}: \emph{method} for eliminating remaining undirected edges.
	\end{itemize}
	\item
	\emph{\textbf{Sample generator}}: matrix generation module, to be used in the test module. It
	generates matrices of three types: Guttman scale, randomly perturbed Guttman scale, 
	and double monotony Mokken scale. In all these cases, the underlying graph is a linear causal chain.
	
	\item
	\emph{\textbf{Test}}: module to evaluate the correctness of the algorithm in simple, self-generated test
	cases, in which the resulting graph is known \emph{a priori}. It generates two types of matrices,
	from perturbations on a base matrix in which, for every component $m_{ij}$, it is true that if
	 $m_{ij} = 1$ then $m_{i,j-1} = 1$ and $m_{i+1,j} = 1$. The perturbations are They are carried out
	completely randomly for matrices of the first type, and with heuristic criteria for those of
	the second type. On each of these matrices, a graph is built and simplified with the
	proposed algorithm and the results obtained are evaluated, depending on the type of
	matrix. The \emph{\textbf{Test}} module receives a set of its own parameters: 
	
	\begin{itemize}
		\item
		\texttt{test\_rows}: number of rows of the test matrix to generate.
		\item
		\texttt{test\_columns}: number of columns of the test matrix to generate.
		\item
		\texttt{test\_perturbation}: probability of making a change (from 0 to 1, or from 1 to 0) 
		in a component of the generated matrix. It is only applicable in the construction of matrices 
		of the first type. 
		\item
		\texttt{test\_cases}: number of test cases.
		\item
		\texttt{magnitudes\_to\_check\_out}: file with monitoring magnitudes for each test case. In
		particular, the fraction of arcs of the underlying graph that were reproduced, the
		fraction of redundant arcs that were eliminated, and the Mokken scalability
		coefficient\brackcite{Mokken} for the test matrix and its transpose are evaluated. 
	\end{itemize}
\end{itemize}

\newpage

The structure of modules and dependencies of the program responds to the following diagram:

\begin{figure}[ht]
	\centering
	\includegraphics[height=10cm]{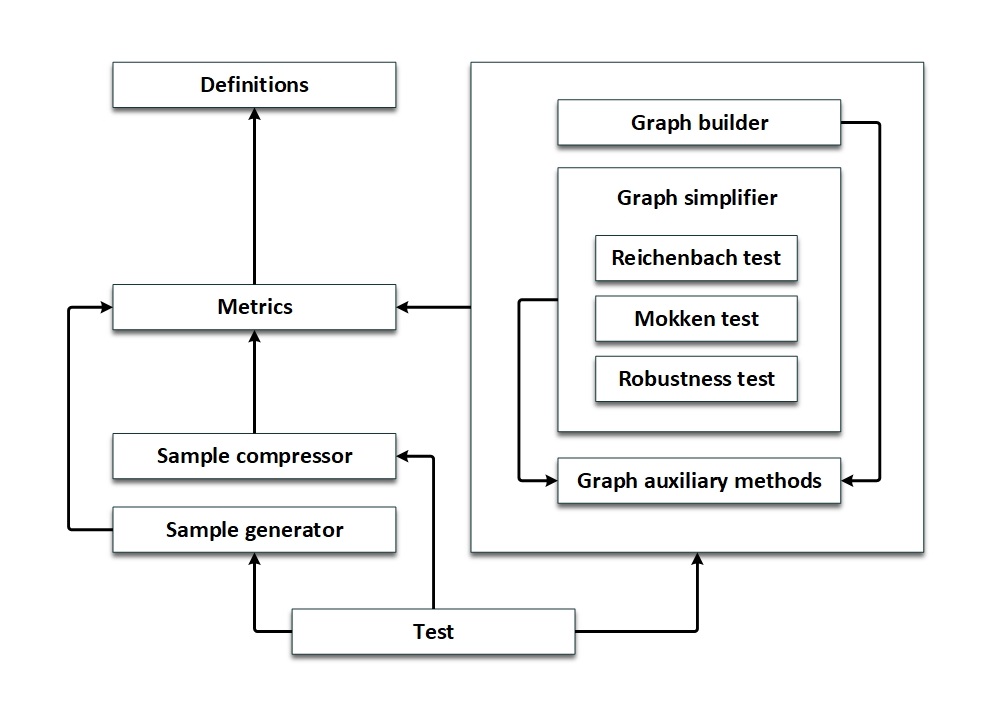}
	\caption{Diagram of dependencies between program modules}
\end{figure}

\section{Implementation details}

In this section, the algorithm used is introduced in general terms and its performance is analyzed.

\subsection{Construction}

The graph is constructed as a list of adjacencies, since the graph is expected to be sparse
(\(|E| < |V|\log|V|\)) or, at least, to be sparse after the simplification process. Although throughout
the algorithm it will continually be necessary to determine the existence of specific arcs within
the graph, for which an adjacency matrix is ideal, it is expected to work with a large number
of variables ($|V|\sim 10^4$), so represent \(G\) as an adjacency matrix of size \(|V| \times |V|\) could
compromise the performance of the algorithm in terms of memory.

To construct the graph, it is necessary to calculate the value $H_{ij}$ for all pairs of
vertices of $G$. Therefore, the frequencies $\pi_i$ are calculated for all vertices of the graph
and stored in a list $\pi$, available in several phases of the algorithm.
Furthermore, the coincidence frequencies $\pi_{ij}$ are calculated for each pair of vertices
but only those corresponding to arcs of $G$ are stored (ie, when $H_{ij}>H_0$). Therefore, in
the adjacency list of $i$, each arc (or undirected edge) $i \rightarrow j$ is represented as a pair $<j, \pi_{ij}>$. 
The structure of the adjacency lists is similar to that which is usual in a weighted graph, although $G$ is not. 
In any case, the analogue of a weight in the graph $G$ is not $\pi_{ij}$ but $H_{ij}~\Theta(H_{ij}-H_0)$.

A function $H(\pi_{ij}, \pi_i, \pi_j)$ is defined that calculates the value $H_{ij}$ for a pair of vertices $i$ and
$j$ of $G$, and an arc $i \rightarrow j$ is added to $G$ if the pair $(i, j)$ satisfies  $\pi_i \leq \pi_j$ and $H_{ij} > H_0$. 
Note that for pairs $(i, j)$, such that $\pi_i = \pi_j$ and $H_{ij} > H_0$, an arc is added in both directions (undirected edges).

\subsubsection{Pseudocode and time complexity}

The following pseudocode reflects the algorithm for constructing the graph. The input parameter $M$ is the base matrix, $\pi$ is the list of frequencies, and $H_0$ is the threshold that the metric must exceed to determine the presence of an arc between two vertices.  

\begin{algorithm}
	\caption{Graph construction algorithm}
	\KwData{$M, \pi, H_0$}
	\KwResult{$G$}
	\SetKwFunction{Fun}{\textbf{Build}}
	\SetKwProg{Fn}{Function}{:}{}
	\Fn{\Fun{$M, \pi, H_0$}}{
			
		$G \gets \text{empty graph}$
		
		\For{$i, j$ \textbf{in} $V \times V$}{
			
			$\pi_{ij} \gets$ \textbf{calculate\_pij}(M, i, j)
			
			\If{$i \neq j$ \textbf{and} $\pi_i \leq \pi_j$ \textbf{and} $H(\pi_{ij}, \pi[i], \pi[j]) > H_0)$}{
				
				\textbf{add} $i \rightarrow j, \pi_{ij}$ \textbf{to} $G$
			}
		}
		\Return $G$
	}
\end{algorithm}

The function \textbf{calcula\_pij} calculates the frequencies of coincidences $\pi_{ij}$ on the
matrix $M$, in a number of operations that linearly depends on the number of individuals.
Therefore, it is $O(|I|)$. This operation is performed once for each pair of vertices. Since
they are a total of $|V|(|V| - 1)$ ordered pairs, and the rest of the operations are atomic,
then the construction method has a time complexity of $O(|I||V|^2)$.

\subsection{Spurious arcs}

The \textbf{Reichenbach\_Test} \emph{method} implements the spurious arc removal algorithm
proposed in \ref{Reichenbach}. For each arc \(i \rightarrow j\) of $G$, the parents common 
to $i$ and $j$ are searched. Of these, we look for some parent $k$ that meets the conditions 
\(H_{ij | k} \leq H_0\) and \(H_{ij | \neg k} \leq H_0\). If this is the case, \(i \rightarrow j\) 
is eliminated from $G$.

For convenience, the graph is transposed, so that each adjacency list of a vertex
contains all its parents instead of all its children. This decision makes it easier to find
the parents common to the ends of an arc \(i \rightarrow j\), finding the intersection of the adjacency
lists of $i$ and $j$.

In accordance with the idea proposed in \ref{Reichenbach_order}, the arcs are ordered by the
average frequency of their ends, and are visited in this order. As seen therein, this
resolves all conflicts between two triangles with shared sides, except when the second
leg of one of the triangles is the hypotenuse of the other, and the second legs involved
are of equal length. In this case, the arcs to visit in each triangle have the same
average frequency and the order criterion is not valid. In this case, the arc with the highest 
correlation between its extremes is taken first. Since conflicting arcs always share a vertex, 
and non-sharing vertices have equal frequencies, for example \(j \rightarrow k\) and \(l \rightarrow k\) 
in \textbf{Fig.}\ref{ReichenbachConflict2}(a), then the ordering of arcs from highest to lowest 
correlation between vertices is reduced to an ordering from highest to lowest frequency of coincidences

\subsubsection{Pseudocode and time complexity}

The following pseudocode reflects the algorithm for removing spurious arcs. The parameters $M$, $\pi$, and $H_0$ are the same as in the previous algorithm, while G is the graph to be simplified.

\begin{algorithm}
	\caption{Algorithm for the elimination of spurious arcs}
	\KwData{$G, M, \pi, H_0$}
	\KwResult{$G$}
	\SetKwFunction{Fun}{\textbf{Reichenbach's\_Test}}
	\SetKwProg{Fn}{Function}{:}{}
	\Fn{\Fun{$G, M, \pi, H_0$}}{
			
		$E \gets E(G)$
		
		\textbf{Sort}$(E,$ \textbf{Compare} $)$
		
		\For{$i \rightarrow j$ \textbf{in} $E$}{
			
			\For{$k$ \textbf{in common\_parents}$(i, j)$}{
				
				\If{\textbf{Common\_cause\_principle}$(M, \pi, H_0, i \rightarrow j, k)$}{
					
					\textbf{remove} $i \rightarrow j, \pi_{ij}$ \textbf{of} $G$
				}
			}
		}
	}

	\SetKwFunction{Func}{\textbf{Compare}}
	\Fn{\Func{$i \rightarrow j, l \rightarrow k$}}{
		
		\If{$\pi[i] + \pi[j] \neq \pi[l] + \pi[k]$}{
			
			\Return $\pi[i] + \pi[j] < \pi[l] + \pi[k]$
		}
		\Return $\pi_{ij} > \pi_{lk}$
	}	 
\end{algorithm}   

\newpage

Where \textbf{Common\_cause\_principle} checks compliance with the conditions 
\(H_{ij | k} \leq H_0\) and \(H_{ij | \neg k} \leq H_0\), calculating \(H_{ij | k}\) y
\(H_{ij | \neg k}\) in \(O(|I|)\).

Ordering the arcs according to the proposed criterion is done in \(O(|E|log|E|) = O(|E|log|V|)\).

Subsequently, for each arc \(i \rightarrow j\) of \(E\), the common parents of \(i\) and \(j\) in \(O(|V|)\) 
(intersection of the adjacency lists of \(i\) and \(j\)) are computed. Then, for each one, the conditions proposed in \(O(|I|)\) are checked. If the conditions are met, the arc is eliminated in \(O(|V|)\) (although it may be reduced to \(O(|log|V|)\) if each adjacency list is implemented over an ordered tree structure or a dictionary). Like one elimination operation is performed for each arc of at most \(O(\log|V|)\), these operations are in total at most at most, these operations are in total \(O(|E||V|)\).

Finally, the time complexity of the spurious arc removal algorithm is 
\(O(|E|log|V| + |E||V||I|) = O(|E||V||I|)\).

\subsection{Redundant arcs}

The \textbf{Mokken\_Test} \emph{method} implements the proposed transitive arc elimination algorithm. For each arc \(i \rightarrow j\) of $G$, it finds all triangles \((i, k, j)\) in which it is hypotenuse and check the conditions \(\pi_{ij} \leq \pi_{ik} \leq \pi_{jk}\) and \(\pi_{\neg i \neg j} \geq \pi_{\neg i \neg k} \geq \pi_{\neg j \neg k}\). If this is fulfilled, then \(i \rightarrow j\) is eliminated from \(G\). Previously, the arcs  are ordered as proposed in the section \ref{TransitiveOrder}, by length from longest to shortest and, in length from longest to shortest, and, in case of equal length, from longest to shortest by  frequency of coincidences between their ends.

\subsection{Pseudocode and time complexity}

The following pseudocode reflects the algorithm for eliminating redundancy arcs. The parameters $G$ and $\pi$ are the same as in the previous algorithm. 

\begin{algorithm}
	\caption{Algorithm for the elimination of redundant arcs}
	\KwData{$G, \pi$}
	\KwResult{$G$}
	\SetKwFunction{Fun}{\textbf{Mokken\_Test}}
	\SetKwProg{Fn}{Function}{:}{}
	\Fn{\Fun{$G, \pi$}}{
		
		$E \gets E(G)$
		
		\textbf{Sort}$(E,$ \textbf{Compare} $)$
		
		\For{$i \rightarrow j$ \textbf{en} $E$}{
				\textbf{Simplify\_triangle}$(G, i \rightarrow j)$	
			}
		}

	\SetKwFunction{Funt}{\textbf{Simplify\_triangle}}
	\Fn{\Funt{$G, i \rightarrow j$}}{
		
		\For{$k$ \textbf{in children}$(i)$}{
			
			\If{\textbf{exists} $j \rightarrow k$ \textbf{in} $G$}{
				\If{\textbf{Mokken\_conditions}$(G, i \rightarrow j)$}{
					\textbf{remove} $i \rightarrow j, \pi_{ij}$ \textbf{of} $G$
					\Return
				}
			}	
		}
	}
	
	\SetKwFunction{Func}{\textbf{Compare}}
	\Fn{\Func{$i \rightarrow j, l \rightarrow k$}}{
		
		\If{$\mid \pi[i] - \pi[j] \mid$ $\neq$ $\mid \pi[l] - \pi[k] \mid$}{
			
			\Return $\mid \pi[i] - \pi[j] \mid$ $>$ $\mid \pi[l] - \pi[k] \mid$
		}
		\Return $\pi_{ij} > \pi_{lk}$
	}	 
\end{algorithm}  

Where \textbf{Mokken\_Conditions} checks the fulfillment of the conditions \(\pi_{ij} \leq \pi_{ik} \leq \pi_{jk}\) 
and \(\pi_{\neg i \neg j} \geq \pi_{\neg i \neg k} \geq \pi_{\neg j \neg k}\),
in \(O(1)\). This is possible since \(\pi_{ij}\), \(\pi_{ik}\), and
\(\pi_{jk}\) are stored in the corresponding arcs, and
\(\pi_{\neg i \neg j}\), \(\pi_{\neg i \neg k}\), y
\(\pi_{\neg j \neg k}\) can be computed from these and the list
\(\pi\) (e.g., \(\pi_{\neg i \neg j} = 1 - \pi_i - \pi_j + \pi_{ij}\)).

Ordering the arcs according to the proposed criterion is done in \(O(|E|log|E|) = O(|E|log|V|)\).

Subsequently, for each arc \(i \rightarrow j\) de \(E\), all the triangles where it is hypotenuse are visited
searching for all vertices $k$ children of $i$ that are parents of $j$, in \(O(|V|^2)\)
(although it can be reduced to \(O(|V|\log|V|)\) by implementing the adjacency lists over 
ordered tree structures or dictionaries). Then, for each one, the conditions proposed in \(O(1)\) are checked .

Finally, the time complexity of the spurious arc elimination algorithm is \(O(|E|log|V| + |E||V|^2) = O(|E||V|^2)\).

\subsection{Remaining undirected edges}

The \textbf{Robustness\_Test} \emph{method} implements the proposed undirected edge removal
algorithm. For each undirected edge \(i \leftrightarrow j\), find the vertices \(i-1\) and \(j-1\), to calculate
the corresponding \(\pi_{ii}\) and \(\pi_{jj}\) values .

\subsubsection{Pseudocode and time complexity}

The following pseudocode reflects the algorithm for orienting or removing remaining undirected edges. The parameters $G$ and $\pi$ are the same as in the previous algorithm.

\begin{algorithm}
	\caption{Algorithm for orienting or eliminating undirected edges}
	\KwData{$G, \pi, H_0$}
	\KwResult{$G$}
	\SetKwFunction{Fun}{\textbf{Robustness\_Test}}
	\SetKwProg{Fn}{Function}{:}{}
	\Fn{\Fun{$G, \pi, H_0$}}{
		
		\For{$i \rightarrow j$ \textbf{in} $E$}{
			
				\Switch{\textbf{Verify\_robustness}$(G, i \leftrightarrow j, M, \pi, H_0)$}{
					
					\Case{$1$}{\textbf{remove} $i \rightarrow j, \pi_{ij}$ \textbf{of} $G$} 

					\Case{$2$}{\textbf{remove} $i \leftarrow j, \pi_{ij}$ \textbf{of} $G$} 
					
					\Case{$3$}{\textbf{remove} $i \leftrightarrow j, \pi_{ij}$ \textbf{of} $G$} 
			}
		}
	}
	
	\SetKwFunction{Funt}{\textbf{Verify\_robustness}}
	\Fn{\Funt{$G, i \leftrightarrow j, \pi, H_0$}}{
		
		$\pi_{ii} \gets$ \textbf{Calculate\_pii}$(G, i \rightarrow j, \pi, H_0)$
		
		$\pi_{jj} \gets$ \textbf{Calculate\_pii}$G, i \leftarrow j, \pi, H_0$
		
		\If{$\pi_{i,i-1} == -\infty$ \textbf{or} $\pi_{j,j-1} == -\infty$}{
			\Return $0$
		}
		
		\Return \textbf{Robustness\_conditions}$(\pi[i], \pi[j], \pi_{ii}, \pi_{jj}, \pi_{ij})$
	}	 

	\SetKwFunction{Funi}{\textbf{Calculate\_pii}}
	\Fn{\Funi{$G, i \rightarrow j, \pi, H_0$}}{
		
		$\pi_{ii} \gets -\infty$; $\pi_{i,i-1} \gets -\infty$; $\pi_{i-1} \gets -\infty$
		
		\For{$k$ \textbf{in parents}$(i)$}{
			
			\If{$k \neq j$}{
				
				\eIf{$\pi_{i,i-1} < \pi_{i,k}$}{
					
					$\pi_{i-1} = \pi[k]$
					
					$\pi_{i,i-1} = \pi_{ik}$
					
					$\pi_{ii} = \pi[i] * \pi{i,i-1} / \pi_{i-1}$
					
				}{
					\If{$\pi_{i,i-1} == \pi_{ik}$ \textbf{and} $\pi_{i-1} < \pi[k]$}{
						
						$\pi_{i-1} = \pi[k]$
						
						$\pi_{ii} = \pi[i] * \pi_{i,i-1} / \pi[k]$
					}
				}
			}
		}
	
		\Return $\pi_{ii}$
	}
\end{algorithm}

\newpage

Where \textbf{Robustness\_Conditions} checks compliance with the robustness conditions
proposed in section \ref{Robustness} by performing the necessary algebraic calculations, in \(O(1)\). 
Returns an integer value indicating the arc that should be removed (eg, \(i \rightarrow j\),
\(i \leftarrow j\) or the entire edge \(i \leftrightarrow j\)).

The time complexity depends on the operations to find \(i-1\) and \(j-1\), which is performed once
for each undirected edge, and is bounded by \(|V|\). Since the number of undirected edges is at
most \(|E|\), then the method is \(O(|V||E|)\).

Consequently, the time complexity of the CChains algorithm is the sum of the time complexities of each of its 
phases, i.e., \(O(|I||V|^2)+O(|E||V||I|)+ O(|E||V|^2)+O(|V||E|)\), which, in the expected case \(|V| \leq |E|\) 
is reduced to \\ \(O(\max(|E||V||I|, |E||V|^2))\).

\chapter{Results}\label{chapter:results}

\subsection{Problem}

Nowadays, scientific-technical developments make it possible to collect large amounts of data in different contexts. In particular, in the field of biology, the human genome project (active from 1990 to 2003) recorded the complete genetic code of man.\brackcite{PGH} Numerous databases are publicly available for researchers in the field to consult in order to find the mechanisms at work in certain phenomena, as well as to test models and theories that might explain them. Many of these repositories are associated with intrinsically causal processes. For example, in the case of the database \textit{The Cancer Genome Atlas}, an enormous amount of information is stored on mutation and gene expression profiles that could reveal the genetic origin of cancer, and indicate strategies to detect and treat it.\brackcite{TCGA} In the latter case, although significant progress has been made in describing groups of cancer-correlated genes, oncogenes and tumor suppressor genes, detailed mechanistic or causal information on carcinogenesis is still lacking. 

Genes are segments of the DNA chain that contain the information necessary for the synthesis of functional molecules (gene products) that perform some function in the cellular environment. The central dogma of molecular biology postulates that this information flows from DNA genes to RNA (gene transcription), and from the latter to proteins (translation), through the mechanism of gene expression. Highly complex processes are involved in the expression of a specific gene, and gene products are involved in different ways. For example, a protein resulting from a translation process may constitute a subunit of the transcription machinery of another gene. For example, RNA polymerases are enzymes of a protein nature that perform essential functions of this mechanism, such as recognizing and binding to specific locations on the DNA molecule to start the transcription process of the corresponding gene. Similarly, the translation process of the RNA chain corresponding to a gene is determined by the results of the gene expression of others. For example, ribosomes are the cytoplasmic organelles where the translation process takes place, and are largely composed of ribosomal RNA, which in turn is the result of transcription processes in certain non-coding genes (those where the transcribed RNA strand is the final gene product and is not translated). Therefore, the results of transcription and translation processes of some genes regulate (or deregulate) the form and quantity in which other gene products are obtained, i.e., the genetic expression of some genes has a causal effect on the genetic expression of others.

These types of causal relationships between genes are known to exist. However, it is not known exactly in specific types of cancer what cause-effect relationships are evident between the genetic dysregulations that trigger carcinogenesis. The task to be developed in this chapter refers to the identification of these operative relationships in cancer, with the aim of detecting genes that play a fundamental role in the regulation of this disease. In the particular case of the present study, prostate cancer is taken as a test case, due to its importance as one of the major causes of death in men worldwide.

\subsection{Data origin}

Gene expression data are obtained from \textit{The Cancer Genome Atlas Program} (TCGA), 
obtained from studies of hundreds of normal and tumor tissue samples,
classified by histopathological techniques.\brackcite{ICIMAF} In the particular case 
of prostate cancer, a total of 551 samples are available, of which 52 correspond to normal 
samples and 499 tumor samples. For each sample, a total of 60,483 genes are recorded 
(the same for all samples). Of these, a subset of 52,870 genes is selected, discarding genes
with null or almost null expression in the study tissue. Expression profiles are
measured through the RNA-Seq technique, and their values are reported in units of
Fragments Per Kilobase of transcription per Million mapped reads (FPKM).\brackcite{ICIMAF}

\subsection{Processing}
\label{Procesing}
Gene expression has a heavy-tailed distribution, with a large number of lowfrequency outliers. 
Therefore, the geometric mean and not the arithmetic mean is selected as the measure to calculate the average gene expression\brackcite{ICIMAF}. Since RNASeq does not accurately detect low expression values, a harmless offset of 0.1 FPKM
is applied to all data\brackcite{ICIMAF}.

Let \(e_{gs}\) be the expression level of gene $g$ in sample $s$, the homeostatic (or reference) gene
expression level is estimated as the geometric mean of \(e_{gs}\) over the set of normal samples $N$, ie, \brackcite{ICIMAF}

\begin{equation}
e_g^{(ref)} = \root_{\mid N \mid}\of{\prod_{s\in N}e_{gs}}
\label{mean}
\end{equation}
where \(|N|\)  is the number of normal samples.

The logarithmic \textit{fold-change} ratio with respect to the reference value is calculated for each sample.

\begin{equation}
\hat e_{gs} = \log_2(\frac{e_{gs}}{e_g^{(ref)}})
\label{foldchange}
\end{equation}

As a result, it is obtained that the over- and under-expression are treated symmetrically. Finally,
the measurement obtained is discretized, considering that a gene \(g\) is overexpressed in a sample \(s\) 
if \(\hat e_{gs} < -1\) (that is, \(e_{gs} < \frac{1}{2}e_g^{(ref)}\)) and underexpressed if 
\(\hat e_{gs} > -1\) (that is, \(e_{gs} > 2e_g^{(ref)}\)). Then, a gene
\(g\)  is altered in a sample \(s\) if \(|\hat e_{gs}| > 1\).

Thanks to discretization, the alteration of a specific gene can be taken as a binary variable, and the study samples as the set of individuals with which it is associated. Then, the matrix \(M\) is constructed so that \(m_{gs} = 1\) if the gene is altered in the sample (\(|hat e_{gs}| > 1\)) and \(m_{gs} = 0\) otherwise (\(|hat e_{gs}| \leq 1\)).

This matrix \(M\) describes the distribution of genetic alterations in the samples, and will be the input of the \emph{CChains} algorithm for the construction of the associated causal graph of genetic alterations. 

\subsection{Artificial cancer state gene}

After it is constructed, an additional row is added to the matrix $M$ , corresponding to an artificial
gene. The variable associated with this gene will have a value of $1$ in all tumor samples (499 samples)
and $0$ in the remaining ones, so that this gene only appears altered in samples of cells that have
entered carcinogenesis. The objective of this false gene is, therefore, to symbolize the cellular cancer
state within the genetic network of deregulations. In this way, a path in the graph that ends in the
artificial gene can be interpreted as a chain of genetic alterations that lead to cancer.

\subsection{Construction of the genetic alterations graph}
From the processed genetic alteration data, the corresponding causal graph of genetic alterations
is constructed, using the \emph{CChains} algorithm. In this graph, an arc \(i \rightarrow j\) represents the cause-effect relationship between the alteration of a gene $i$ and that of another gene $j$, while a path represents a causal chain of said alterations.

\subsection{Results}

The constructed matrix $M$ has a total of 551 columns (one for each tissue sample in the data) and
52,870 rows (one for each gene recorded in the samples). This is compressed as proposed in \ref{Compress},
and an equivalent matrix of 46923 rows is obtained, where the identical variables in the original matrix
start to occupy a single row.
The new matrix $M$ is used as input to the algorithm for the construction of the causal graph $G$.

The characteristics of $G$, after each phase of the algorithm, are the following:

\begin{itemize}
\item
Construction of the causal graph:

\begin{itemize}
	\item
	46 923 vertices
	\item
	69 583 711 arcs
	\item
	132 755 undirected edges
\end{itemize}
\item
Removal of spurious arcs:

\begin{itemize}
	\item
	22 357 509 arcs
	\item
	47 undirected edges
\end{itemize}

Therefore, 47226202 arcs and 132708 undirected edges are eliminated.
\item
Elimination of redundant arcs:

\begin{itemize}
	\item
	10 664 000 arcs
	\item
	No undirected edges
\end{itemize}

Therefore, 11693556 arcs are eliminated, non-directed edges are not eliminated, and 47 of
these are oriented.
\item
Removal of undirected edges:

\begin{itemize}
	\item
	Since the graph does not have undirected edges upon reaching this phase, no change occurs.
\end{itemize}
\end{itemize}

Therefore, a graph $G$ with 46,923 vertices and 10,664,000 arcs is obtained, without
undirected edges. It is also true that it is acyclic, a fact that was verified by calculating the
topological order of $G$ using the Kahn algorithm.\brackcite{Kahn} Furthermore, the undirected graph
underlying $G$ consists of a single connected component, that is, in $G$ there are no isolated
vertices.
The distribution of degrees in graph $G$ is quite irregular, and is described by the following graph:

\begin{figure}[ht]
	\centering
	\includegraphics[height=9cm]{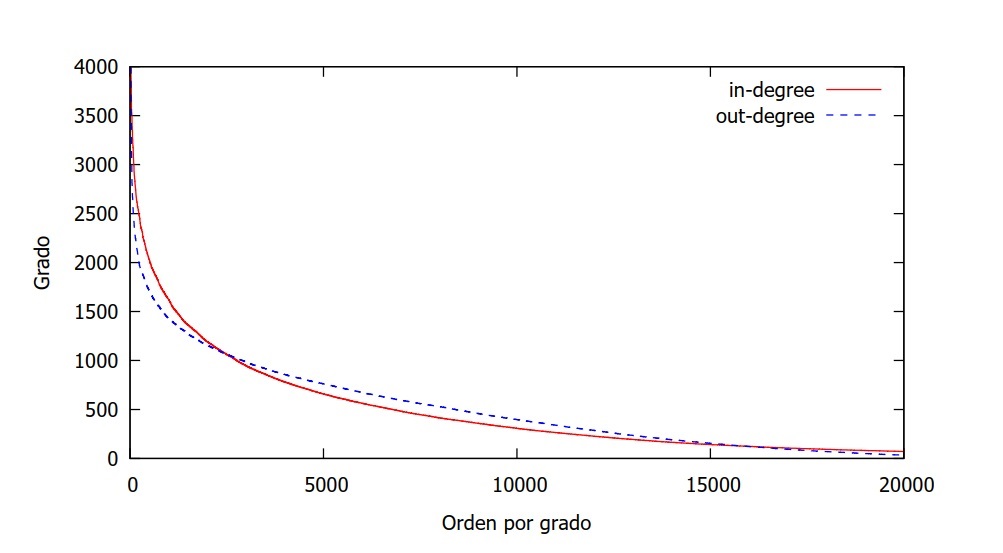}
	\caption{Distribution of degrees in the graph of genetic alterations}
	\label{Contrast}
\end{figure}

We also have that 467 vertices are orphans, and 19460 are childless vertices. Of these 467, 420 have paths that lead to the artificial vertex of cancer.

A large number of vertices are concentrated around the artificial cancer vertex. A total of 3872 vertices have this vertex as a child, 12767 are ancestors at distance 2, and about 13105 at distance 3.

To illustrate the above, see the corresponding \textbf{Fig.}\ref{Contrast} a subgraph of radius 2, around the fictitious cancer vertex, in which only vertices of \emph{contrast} greater than 5\% (at least 5\% of their adjacencies are the cancer vertex or one of its ancestors) are considered. The vertices correspond to the genes in the \ref{Contrast1} and \ref{Contrast2} tables, in order of numbering. The genes are accompanied by their Ensembl IDs, which are the stable identifiers by which they can be located in the Ensembl database.\brackcite{Ensembl}  

\begin{figure}[ht]
	\centering
	\includegraphics[height=9cm]{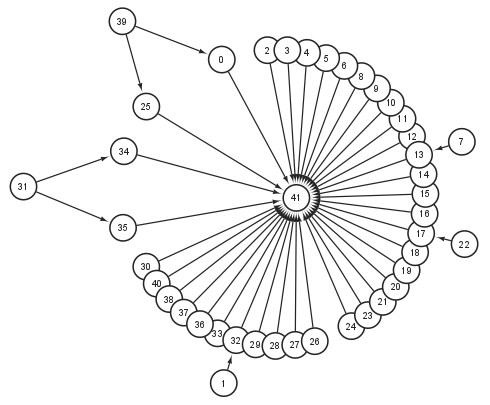}
	\caption{Subgraph of the simplified graph, induced by the vertices with contrast greater than 5\%}
\end{figure}

\begin{table}[H]
	
	\centering	
	\resizebox{250pt}{150pt}{%
		
		\begin{tabular}{ccc}
			\toprule
			Número & Gen & Ensembl ID \\
			\midrule
			\midrule
			0 & MIR1299 & ENSG00000275377 \\
			1 & - & ENSG00000223180 \\
			2 & - & ENSG00000221211 \\
			3 & RNA5SP199 & ENSG00000200275 \\
			4 & RNU6-32P & ENSG00000206675 \\
			5 & - & ENSG00000280673 \\
			6 & AC087393.1 & ENSG00000263729 \\
			7 & RNU6-1107P & ENSG00000201687 \\
			8 & - & ENSG00000263913 \\
			9 & Y\_RNA & ENSG00000252894 \\
			10 & HP & ENSG00000257017 \\
			11 & RPL35P6 & ENSG00000244018 \\
			12 & SETD6P1 & ENSG00000236877 \\
			13 & AC006463.1 & ENSG00000225795 \\
			14 & AC023421.1 & ENSG00000266968 \\
			15 & AC024257.2 & ENSG00000258273 \\
			16 & LINC02244 & ENSG00000259590 \\
			17 & AC104985.1 & ENSG00000267746 \\
			18 & TRAJ45 & ENSG00000211844 \\
			19 & AC092653.1 & ENSG00000273245 \\
			20 & Y\_RNA & ENSG00000206817 \\
			\bottomrule	
		\end{tabular}%
	}	
	\caption{Vertices with contrast greater than 5\%}
	\label{Contrast1}
\end{table}			

\begin{table}[H]
	
	\centering	
	\resizebox{250pt}{150pt}{%
		
		\begin{tabular}{ccc}
			\toprule
			Número & Gen & Ensembl ID \\
			\midrule
			\midrule
			21 & LINC00533 & ENSG00000235570 \\
			22 & AC006249.1 & ENSG00000274578 \\
			23 & Y\_RNA & ENSG00000207480 \\
			24 & RNA5SP452 & ENSG00000199874 \\
			25 & - & ENSG00000273631 \\
			26 & Y\_RNA & ENSG00000199979 \\
			27 & RNU6-514P & ENSG00000206935 \\
			28 & Y\_RNA & ENSG00000200118 \\
			29 & - & ENSG00000277347 \\
			30 & RNU6-906P & ENSG00000207431 \\
			31 & ATG3 & ENSG00000144848 \\
			32 & RNU6-575P & ENSG00000223258 \\
			33 & AP005136.1 & ENSG00000238575 \\
			34 & AP000350.5 & ENSG00000272973 \\
			35 & AC003072.1 & ENSG00000250318 \\
			36 & RNU6-858P & ENSG00000199306 \\
			37 & TRDJ2 & ENSG00000211827 \\
			38 & - & ENSG00000216067 \\
			39 & AL360157.1 & ENSG00000260574 \\
			40 & FAM136BP & ENSG00000232654 \\
			41 & CancerGene & - \\
			\bottomrule
		\end{tabular}%
	}	
	\caption{Vertices with contrast greater than 5\%}
	\label{Contrast2}
\end{table}

\begin{figure}[ht]
	\centering
	\includegraphics[height=9cm]{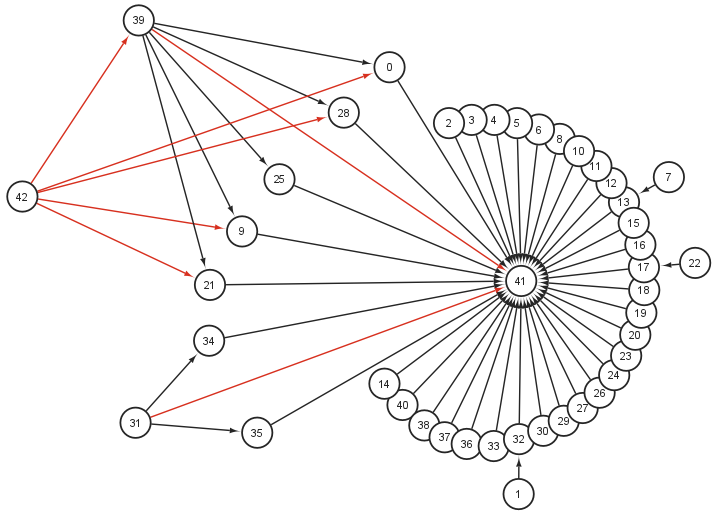}
	\caption{Subgraph of the base network, induced by vertices with contrast greater than 5\% in the simplified network}
	\label{ContrastComplete}
\end{figure}

The subgraph induced by these vertices, prior to the application of the spurious and redundant arcs elimination algorithms, had the structure described in the \textbf{Fig.}\ref{ContrastComplete}. As can be seen, in addition to the above relationships, there were spurious arcs (39 \(39 \rightarrow 9\), \(39 \rightarrow 21\) and \(39 \rightarrow 28\). The vertex \(42\) is added to the subgraph. corresponding to AC008871.1 (ENSG00000250383) is added to the subgraph, since it is the common parent of to \(9\), \(21\), and \(28\), responsible for the appearance of these arcs, eliminated in the corresponding phase. On the other hand, \(42\) is also the parent of vertex \(0\), but the connection \(39 \rightarrow 0\) is not lost, since it is not spurious. Note that there were also present the arcs \(39 \rightarrow 41\) and \(31 \rightarrow 41\) were also present, representing transitive relationships, and were therefore eliminated. On the other hand, the phenomenon of \emph{loss of transitivity} explained in the section \ref{NonTransitivity}, in the causal chains \(1 \rightarrow 32 \rightarrow 41\), \(22 \rightarrow 17 \rightarrow 41\) and \(7 \rightarrow 13 \rightarrow 41\) (the arcs \(1 \rightarrow 41\), \(22 \rightarrow 41\) and \(7 \rightarrow 41\) do not exist).

\newpage
To measure the importance of each gene in the deregulation network, a PageRank
algorithm is used. PageRank is a family of algorithms created and developed by the Google
company to classify web pages by importance and optimize their search engines. It is based
on the initial idea that the ``importance'' of a page depends in turn on that of all the pages that
have links to it, and measures this value with a measure homonymous to the algorithm. Thus,
the PageRank of a page defines its importance on the network. In the analysis proposed in
this study, this notion is used in reverse: the importance of a vertex in the graph depends on
the importance of all its child vertices, ie, the PageRank of a vertex under this new definition
is the PageRank of the vertex in the transposed graph, using the usual definition. The idea
with this is to qualify the deregulatory potential of each gene based on the deregulatory
potential of all the genes altered by it. The PageRank of the vertices of the transposed graph
is then calculated and ordered by it, from highest to lowest, to obtain a ranking by importance.
The first 15 vertices of the ranking correspond to table \ref{Pagerank}, which shows the main data of
each vertex, that is, index of the vertex in the resulting graph, number of genes corresponding
to said vertex (due to the compression process), number of arcs from the vertex to others
(\textit{out-degree}), number of arcs from other vertices to this (\textit{in-degree}) 
and PageRank calculated on it. 

\begin{table}[H]
	
	\centering	
	\resizebox{450pt}{130pt}{%
		
		\begin{tabular}{ccccc}
		\toprule
				Vértice & Cantidad de genes & Out-degree & In-degree & Ranking \\
				\midrule
				\midrule
				106 & 904 & 296 & 0 & 0.0280089084549199\\
				203 & 387 & 231 & 0 & 0.0169252646143832\\
				53 & 266 & 194 & 0 & 0.0138095007882416\\
				232 & 405 & 4631 & 2 & 0.00951893530407698\\
				54 & 200 & 4037 & 2 & 0.0070904279384024\\
				894 & 67 & 3647 & 2 & 0.00593953728884232\\
				107 & 135 & 186 & 0 & 0.00483720906329685\\
				59 & 108 & 126 & 0 & 0.00346518187070434\\
				1518 & 102 & 112 & 0 & 0.00314751044972811\\
				477 & 70 & 144 & 0 & 0.00291192330216641\\
				126 & 22 & 219 & 0 & 0.0026924875847944\\
				1476 & 58 & 119 & 0 & 0.0024063629481943\\
				1810 & 78 & 172 & 0 & 0.00227731537509089\\
				669 & 52 & 114 & 0 & 0.00221713079450858\\
				2504 & 16 & 13469 & 0 & 0.00218459734706735\\
			\bottomrule
		\end{tabular}%
		}	
		\caption{First 15 vertices of the ranking by PageRank}
		\label{Pagerank}
\end{table}

The PageRank algorithm evaluates the significance of each vertex in the network considering the entire network. However, to analyze the significance of each vertex with respect to cancer, the algorithm is evaluated only on the subgraph of \(G\) which comprises only the vertices that have paths to the cancer vertex, and the connections between these. The first 15 genes correspond to the table \ref{Inducedpagerank}: 

\begin{table}[H]
	
	\centering
	\resizebox{450pt}{120pt}{%
		
		\begin{tabular}{ccccc}
			\toprule
			Vértice & Cantidad de genes & Out-degree & In-degree & Ranking \\
			\midrule
			\midrule
			106 & 904 & 296 & 0 & 0.0386629717239762\\
			203 & 387 & 231 & 0 & 0.0228670690866003\\
			53 & 266 & 194 & 0 & 0.0191475773473782\\
			232 & 405 & 4631 & 2 & 0.0141473304810091\\
			54 & 200 & 4037 & 2 & 0.0107754221143982\\
			894 & 67 & 3647 & 2 & 0.00868884719466811\\
			107 & 135 & 186 & 0 & 0.00713393130155398\\
			1476 & 58 & 119 & 0 & 0.00393586413393418\\
			238 & 38 & 94 & 0 & 0.00373306252059571\\
			1518 & 102 & 112 & 0 & 0.00354108602504106\\
			59 & 108 & 126 & 0 & 0.00351365491373061\\
			1810 & 78 & 172 & 0 & 0.00321007251876712\\
			477 & 70 & 144 & 0 & 0.00310394431725929\\
			126 & 22 & 219 & 0 & 0.00296362179658831\\
			1270 & 47 & 118 & 0 & 0.00293891200199811\\
			\bottomrule
	\end{tabular}%
	}	
	\caption{First 15 vertices of the ranking by PageRank, in the graph induced by vertices with paths to the artificial vertex of cancer}
	\label{Inducedpagerank}
\end{table}

Next, it is of interest to examine the genes that emerge from a principal component analysis (PCA) on the same data used in this study. It is worth recapitulating several notable results of the above PCA: 1) normal and tumor samples are evidently separated on the first principal component, 2) it is possible to establish a hierarchy or \textit{ranking} of genes according to their contribution to the first principal component, and 3) a small number of genes ($32$) from this hierarchy suffices to satisfactorily rank the samples. The hierarchy of genes by PageRank is similar to the one calculated by PCA (2), with the important difference that the PageRank in graded in the \ref{Pagerank} table depends on an asymmetric measure of connection between variables/vertices (Loevinger coefficient) whereas the \textit{ranking} provided by PCA is built on symmetric matrices of correlations or covariances. In fact, a gene at the top of the PageRank hierarchy of a directed network will not necessarily appear prioritized in the \textit{ranking} of genes in the corresponding transposed network or the underlying undirected network. Another significant difference is that the PCA analysis directly employs the fold-change (\ref{foldchange}) as a continuous measure, not discretizing it in the manner proposed in the \ref{Procesing} section. 

The vertices corresponding to result 3) of this PCA analysis occupy a relatively low place in the calculated PageRank, as can be seen in the tables \ref{Pageranka} and \ref{Pagerankb}.        

\begin{table}[H]
	\centering	
	\resizebox{450pt}{130pt}{%
		
		\begin{tabular}{cccccc}
			\toprule
			Gen & Ensembl ID & Vértice &  In-degree & Ranking & Posición en el rank \\
			\midrule
			\midrule
			DLX1 & ENSG00000144355 & 20565 & 3186 & 8.33393955008047e-06 & 31609\\
			PCA3 & ENSG00000225937 & 8137 & 4442 & 8.33393955008047e-06 & 31909\\
			AP006748.1 & ENSG00000223400 & 40665 & 3353 & 8.33393955008047e-06 & 31769\\
			AL359314.1 & ENSG00000274326 & 34974 & 2773 & 8.33393955008047e-06 & 27860\\
			RPL7P16 & ENSG00000242899 & 39644 & 2529 & 8.33393955008047e-06 & 32778\\
			HOXC6 & ENSG00000197757 & 46096 & 3504 & 8.33393955008047e-06 & 32868\\
			ARLNC1 & ENSG00000260896 & 20221 & 2829 & 8.33393955008047e-06 & 32134\\
			PCAT14 & ENSG00000280623 & 15811 & 3196 & 8.33393955008047e-06 & 28480\\
			AP001610.3 & ENSG00000232806 & 555 & 2873 & 8.33393955008047e-06 & 30696\\
			AMACR & ENSG00000242110 & 17232 & 1917 & 8.33393955008047e-06 & 30085\\
			COMP & ENSG00000105664 & 24764 & 2050 & 8.33393955008047e-06 & 33009\\
			SIM2 & ENSG00000159263 & 866 & 4038 & 8.33393955008047e-06 & 31729\\
			AC092535.5 & ENSG00000273179 & 2734 & 2625 & 8.33393955008047e-06 & 34137\\
			TDRD1 & ENSG00000095627 & 4970 & 1529 & 8.33393955008047e-06 & 28812\\
			AP002498.1 & ENSG00000254988 & 1093 & 2148 & 8.33393955008047e-06 & 34382\\
			OR51E2 & ENSG00000167332 & 489 & 2228 & 8.33393955008047e-06 & 35117\\
			HPN & ENSG00000105707 & 4241 & 2463 & 8.33393955008047e-06 & 32708\\
			TRGC1 & ENSG00000211689 & 15036 & 3849 & 8.33393955008047e-06 & 30160\\
			SLC45A2 & ENSG00000164175 & 4870 & 1580 & 8.33393955008047e-06 & 30533\\
			AC139783.1 & ENSG00000250767 & 18423 & 1979 & 8.33393955008047e-06 & 30036\\
			\bottomrule
	\end{tabular}%
	}	
	\caption{PageRank of the vertices corresponding to the genes of the PCA analysis}
	\label{Pageranka}
\end{table} 			
			
\begin{table}[H]
	\centering	
	\resizebox{450pt}{100pt}{%
		
		\begin{tabular}{cccccc}
			\toprule
			Gen & Ensembl ID & Vértice & In-degree & Ranking & Posición en el rank \\
			\midrule
			\midrule			
			HOXC4 & ENSG00000198353 & 24217 & 1996 & 8.33393955008047e-06 & 33013\\
			TRGV9 & ENSG00000211695 & 41519 & 2364 & 8.33393955008047e-06 & 27660\\
			TP63 & ENSG00000073282 & 3558 & 1365 & 8.33393955008047e-06 & 30581\\
			CRTAC1 & ENSG00000095713 & 28554 & 3556 & 8.33393955008047e-06 & 31561\\
			KRT5 & ENSG00000186081 & 29782 & 2317 & 8.33393955008047e-06 & 29638\\
			GPX2 & ENSG00000176153 & 8765 & 1718 & 8.33393955008047e-06 & 37926\\
			WFDC2 & ENSG00000101443 & 22645 & 2113 & 8.33393955008047e-06 & 33020\\
			GSTM1 & ENSG00000134184 & 23068 & 1656 & 8.33393955008047e-06 & 32503\\
			SERPINA5 & ENSG00000188488 & 23401 & 2742 & 8.33393955008047e-06 & 29863\\
			SLC39A2 & ENSG00000165794 & 30024 & 2764 & 8.33393955008047e-06 & 31823\\
			ACTC1 & ENSG00000159251 & 10114 & 3533 & 8.33393955008047e-06 & 31678\\
			SEMG2 & ENSG00000124157 & 41127 & 3543 & 8.33393955008047e-06 & 29250\\
			SEMG1 & ENSG00000124233 & 36756 & 5766 & 8.33393955008047e-06 & 31509\\
			\bottomrule
		\end{tabular}%
	}	
	\caption{PageRank of the vertices corresponding to the genes in the analysis by PCA}
	\label{Pagerankb}
\end{table}   

Each of these vertices corresponds only to the analogous gene (they do not compress other genes), and all are childless vertices (\textit{out-degree=0}). The difference in the results of the two studies may be due to the symmetrical nature of the former. The importance of these genes in the prostate cancer dysregulation network may not be due to their dysregulatory potential in the network, i.e., as important causes of dysregulation. It may be due, instead, to a high probability of being dysregulated when the tissue has begun the process of carcinogenesis, perhaps due to a common cause.
Strong support for the above hypothesis is that in the PageRank calculated on the untransposed network, these genes occupy high positions in the resulting ranking (tables \ref{Pagerankc} and \ref{Pagerankd}). This ranking, as opposed to the previous one, measures the importance of each vertex by that of its parent vertices, that is, the potential of each gene to be deregulated based on the same measure of the genes that deregulate it.

\begin{table}[H]
	\centering	
	\resizebox{450pt}{130pt}{%
		
		\begin{tabular}{cccccc}
			\toprule
			Gen & Ensembl ID & Vértice &  In-degree & Ranking & Posición en el rank \\
			\midrule
			\midrule
			DLX1 & ENSG00000144355 & 20565 & 3186 & 0.000164793805964374 & 151\\
			PCA3 & ENSG00000225937 & 8137 & 4442 & 0.000254304840322668 & 35\\
			AP006748.1 & ENSG00000223400 & 40665 & 3353 & 0.000162914728880933 & 158\\
			AL359314.1 & ENSG00000274326 & 34974 & 2773 & 0.000165346803992387 & 148\\
			RPL7P16 & ENSG00000242899 & 39644 & 2529 & 0.000268863327673889 & 32\\
			HOXC6 & ENSG00000197757 & 46096 & 3504 & 0.000193321361183645 & 82\\
			ARLNC1 & ENSG00000260896 & 20221 & 2829 & 0.000150749974167285 & 189\\
			PCAT14 & ENSG00000280623 & 15811 & 3196 & 0.000150343735332684 & 193\\
			AP001610.3 & ENSG00000232806 & 555 & 2873 & 0.000125864681798513 & 318\\
			AMACR & ENSG00000242110 & 17232 & 1917 & 0.000135374156142189 & 254\\
			COMP & ENSG00000105664 & 24764 & 2050 & 0.000213930674223797 & 61\\
			SIM2 & ENSG00000159263 & 866 & 4038 & 0.00027256701492603 & 31\\
			AC092535.5 & ENSG00000273179 & 2734 & 2625 & 0.000110463834864383 & 459\\
			TDRD1 & ENSG00000095627 & 4970 & 1529 & 7.34848942035608e-05 & 1151\\
			AP002498.1 & ENSG00000254988 & 1093 & 2148 & 9.96802582240097e-05 & 583\\
			OR51E2 & ENSG00000167332 & 489 & 2228 & 0.00014917499465663 & 194\\
			HPN & ENSG00000105707 & 4241 & 2463 & 0.000163927728714361 & 154\\
			TRGC1 & ENSG00000211689 & 15036 & 3849 & 0.000224990760449499 & 48\\
			SLC45A2 & ENSG00000164175 & 4870 & 1580 & 6.26704655930819e-05 & 1577\\
			AC139783.1 & ENSG00000250767 & 18423 & 1979 & 8.33662001639185e-05 & 871\\
			\bottomrule
		\end{tabular}%
	}	
	\caption{PageRank of the vertices corresponding to genes from PCA analysis}
	\label{Pagerankc}
\end{table} 			

\begin{table}[H]
	\centering	
	\resizebox{450pt}{100pt}{%
		
		\begin{tabular}{cccccc}
			\toprule
			Gen & Ensembl ID & Vértice & In-degree & Ranking & Posición en el rank \\
			\midrule
			\midrule			
			HOXC4 & ENSG00000198353 & 24217 & 1996 & 0.000156169111412572 & 173\\
			TRGV9 & ENSG00000211695 & 41519 & 2364 & 0.000106160510958864 & 507\\
			TP63 & ENSG00000073282 & 3558 & 1365 & 4.41089879412608e-05 & 2857\\
			CRTAC1 & ENSG00000095713 & 28554 & 3556 & 0.000217476069832593 & 57\\
			KRT5 & ENSG00000186081 & 29782 & 2317 & 7.00959440693382e-05 & 1289\\
			GPX2 & ENSG00000176153 & 8765 & 1718 & 7.9655960496375e-05 & 964\\
			WFDC2 & ENSG00000101443 & 22645 & 2113 & 7.20540863212926e-05 & 1207\\
			GSTM1 & ENSG00000134184 & 23068 & 1656 & 7.01189032374503e-05 & 1286\\
			SERPINA5 & ENSG00000188488 & 23401 & 2742 & 0.000156867918023487 & 171\\
			SLC39A2 & ENSG00000165794 & 30024 & 2764 & 0.000232173817725135 & 43\\
			ACTC1 & ENSG00000159251 & 10114 & 3533 & 0.000162686091206463 & 159\\
			SEMG2 & ENSG00000124157 & 41127 & 3543 & 0.000289345829676181 & 24\\
			SEMG1 & ENSG00000124233 & 36756 & 5766 & 0.000423225689758056 & 8\\
			\bottomrule
		\end{tabular}%
	}	
	\caption{PageRank of the vertices (without transposition of the graph) corresponding to genes from PCA analysis}
	\label{Pagerankd}
\end{table}

\backmatter

\begin{conclusions}
In the present work, an analysis of some of the theoretical currents that study
causality was carried out, and some of the main concepts and strategies of current
methods of causal discovery were examined. A scheme for the discovery of causal
sufficiency relationships was presented, similar to the probabilistic theory, with several
original ideas. The proposed algorithm shows a different approach for the discovery
of causal relationships, which, with due analysis and development, can represent an
alternative to current methods, mainly in cases of causal modeling on \textit{Big Data}.

A program was designed for scientific use, in the C++ programming language,
which includes the ideas of the algorithm and is currently in use. As a particular
application, a network of genetic deregulations associated with cancer is found and
modeled, in the specific case of prostate adenocarcinoma. This network can be the
starting point for intervention studies, allowing a better understanding of carcinogenesis.	 
\end{conclusions}

\begin{recomendations}
	The methods of the proposed algorithm still need to be analyzed in detail, to obtain
	have a general notion of their behavior in different areas.
	
	On this basis, the following investigations are recommended:
    
    \begin{itemize}
    	\item
		Find and implement a solution to the ordering problems of the spurious and
		redundant arc elimination methods, respectively, for cases where they fail.
    	\item
    	Find all the transitive causal chains of the graph, particularly those that reach the
    	vertex of the cancer. Furthermore, find the causal chains where all arcs \(i \rightarrow j\) 
    	of the associated path have \(H_{ij}=1\).
    	\item
    	Develop gene rankings by other metrics.
    	\item
    	Find minimal gene panels for cancer diagnosis, ie, minimally necessary sets of
    	genes to alter for carcinogenesis to occur.
    	\item
    	Computational improvements to the program.
    	
    	Use dictionaries or ordered tree structures instead of lists. Sort the vertices and
    	their adjacencies by frequency of the associated variable. Implement algorithms
    	designed \textit{ad hoc} to take advantage of this order. Consider parallelization in the
    	phases of the algorithm where possible.
    	\item
    	Study of the convergence of the algorithm with respect to the number of individuals.
    	From a set of data, referring to a set $I$ of individuals and $V$ of associated variables,
    	make a succession of sets of individuals $I_i$ where each one is a superset of the
    	previous one and a subset of $I$. Construct the succession of graphs $G_i$ resulting from
    	executing the algorithm on the set $V$ of variables, for each set $I_i$. Check if the
    	distribution of arcs of the graphs $G_i$ converges to the distribution in the final graph.
    	\item
    	Detailed study of transitivity. Particularly, the possibility of a redundant arc being
    	maintained by being the shared hypotenuse of two transitive triangles. If this is
    	the case, analyze possible solutions to eliminate the redundant arc.
    	\item
    	Incorporate methods based on Hausman's principle of causal independence to
    	guide double arcs, adapting it appropriately for causal networks where transitivity
    	does not necessarily occur.
    	\item
		Check the causal sufficiency of the causal relationships discovered by the
		algorithm, under the proposals of Cartwright or Skyrme, or use Baumgartner's
		scheme to find minimal probabilistic INUS conditions.
    	\item
		Comparison, on the same data sets, of the results of the algorithm with those of
		other causal discovery algorithms, such as the PC algorithm.
    	\item
		Comparison of the graphs constructed by the algorithm with co-expression
		networks modeled on the same data sets.
    	\item
		Study of the behavior of interventions on the causal graphs constructed, using
		Pearl's do-calculus.
    	\item
    	Construction of a graph that represents the causal relationships of prevention.
    	Design of a scheme to eliminate spurious and redundant arcs for the prevention
    	graph.
    	\item
    	Consider modifications to the algorithm for sets of non-binary variables.
    \end{itemize}
    
\end{recomendations}

\printbibliography[heading=bibintoc]

@book{Hume,
  title={Tratado de la naturaleza humana},
  author={Hume, David},
  year={2020},
  publisher={Editorial Verbum}
}

@article{Nobel,
  title={Premio Nobel de Econom{\'i}a 2021 - David Card, Joshua Angrist y Guido Imbens ganan el galard{\'o}n por sus an{\'a}lisis del mercado laboral},
  author={BBC News Mundo},
  journal={BBC News Mundo},
  year={2021}
}

@article{Turing,
  title={Turing Award: Judea Pearl},
  author={J. Russell, Stuart},
  journal={A.M.Turing Award},
  year={2011}
}

@article{Turing2,
  title={Judea Pearl: Ganador del Premio Turing 2012},
  author={Michelone, Manuel},
  journal={UnoCero},
  year={2012}
}

@article{Reg,
  title={Regularity and Inferential Theories of Causation},
  author={Holger, Andreas and Guenther, Mario},
  journal={Stanford Encyclopedia of Philosophy},
  year={2021}
}

@article{War,
  title={Centenario de la Primera Guerra Mundial},
  author={El Portal del Lector},
  journal={El Portal del Lector},
  year={2014}
}

@article{Myopia,
  title={Night Lights don't lead to nearsightedness, study suggest},
  author={Wagner, Holly},
  journal={The Ohio State University, Research News},
  year={2014}
}

@article{Mill,
  title={A system of logic, ratiocinative and inductive: Being a connected view of the principles of evidence, and the methods of scientific investigation, Vol. 1},
  author={Mill, John Stuart},
  year={1875},
  publisher={Longmans, Green, Reader, and Dyer}
}

@article{Mackie,
  title={Causes and conditions},
  author={Mackie, John L},
  journal={American philosophical quarterly},
  volume={2},
  number={4},
  year={1965},
  publisher={JSTOR}
}

@article{Baumgartner,
  title={A regularity theoretic approach to actual causation},
  author={Baumgartner, Michael},
  journal={Erkenntnis},
  volume={78},
  number={1},
  year={2013},
  publisher={Springer}
}

@article{Cartwright,
  title={Causal laws and effective strategies},
  author={Cartwright, Nancy},
  journal={No{\^u}s},
  year={1979},
  publisher={JSTOR}
}

@article{Prob,
  title={Probabilistic Causation},
  author={Hitchcock, Christopher},
  journal={Stanford Encyclopedia of Philosophy},
  year={2018}
}

@book{Reichenbach,
  title={The direction of time},
  author={Reichenbach, Hans},
  volume={65},
  year={1991},
  publisher={Univ of California Press}
}

@book{Hausman,
  title={Causal asymmetries},
  author={Hausman, Daniel M and a Simon, Herbert and others},
  year={1998},
  publisher={Cambridge University Press}
}

@book{Skyrms,
  title={Causal Necessity},
  author={Skyrms, Brian},
  year={1980},
  publisher={New Haven and London: Yale University Press}
}

@article{Mani,
  title={Causation and Manipulability},
  author={Woodward, James},
  journal={Stanford Encyclopedia of Philosophy},
  year={2016}
}

@article{Menzies,
  title={Causation as a secondary quality},
  author={Menzies, Peter and Price, Huw},
  journal={The British Journal for the Philosophy of Science},
  volume={44},
  number={2},
  pages={187--203},
  year={1993},
  publisher={Oxford University Press}
}

@book{Pearl,
  title={Causality},
  author={Pearl, Judea},
  year={2009},
  publisher={Cambridge university press}
}

@article{Lewis,
  title={Philosophical papers volume I},
  author={Lewis, David},
  year={1983}
}

@book{Mellor,
  title={The facts of causation},
  author={Mellor, David Hugh},
  year={2002},
  publisher={Routledge}
}

@book{Neal,
  title={Introduction to Causal Inference, from a Machine Learning Perspective},
  author={Neal, Brady},
  year={2017}
}

@book{Spirtes,
  title={Causation, prediction, and search},
  author={Spirtes, Peter and Glymour, Clark N and Scheines, Richard and Heckerman, David},
  year={2000},
  publisher={MIT press}
}

@incollection{Mokken,
  title={A theory and procedure of scale analysis},
  author={Mokken, Robert Jan},
  booktitle={A theory and procedure of scale analysis},
  year={2011},
  publisher={De Gruyter Mouton}
}

@article{PGH,
  title={Finishing the euchromatic sequence of the human genome},
  author={International Human Genome Sequencing Consortium},
  journal={Nature},
  volume={431},
  number={7011},
  year={2004},
  publisher={Nature Publishing Group UK London}
}

@article{TCGA,
  title={The cancer genome atlas pan-cancer analysis project},
  author={Weinstein, John N and Collisson, Eric A and Mills, Gordon B and Shaw, Kenna R and Ozenberger, Brad A and Ellrott, Kyle and Shmulevich, Ilya and Sander, Chris and Stuart, Joshua M},
  journal={Nature genetics},
  volume={45},
  number={10},
  year={2013},
  publisher={Nature Publishing Group}
}

@article{ICIMAF,
  title={Perfect genetic biomarkers for cancer from a fresh view of gene dysregulation},
  author={Gil, Gabriel and Gonzalez, Augusto},
  journal={Biorxiv},
  year={2022}
}

@article{Kahn,
  title={Topological sorting of large networks},
  author={Kahn, Arthur B},
  journal={Communications of the ACM},
  year={1962},
  publisher={ACM New York, NY, USA}
}

@article{Ensembl,
  title={The Ensembl genome database project},
  author={Hubbard, Tim and Barker, Daniel and Birney, Ewan and Cameron, Graham and Chen, Yuan and Clark, L and Cox, Tony and Cuff, J and Curwen, Val and Down, Thomas and others},
  journal={Nucleic acids research},
  volume={30},
  number={1},
  pages={38--41},
  year={2002},
  publisher={Oxford University Press}
}

\end{document}